\documentclass[a4paper,fleqn,amssymb,usenatbib,useAMS,usedcolumns]{mnras}
\usepackage{times} 
\usepackage{latexsym,amssymb,amsmath,amsfonts}
\usepackage{rotating} 
\usepackage{float}
\usepackage{graphicx}
\usepackage{natbib}
\usepackage{longtable}
\usepackage{url}
\usepackage{dcolumn}
\usepackage{textcomp}
\usepackage{multicol}
\usepackage{bm}
\usepackage{pdflscape}
\usepackage{lscape}
\usepackage[T1]{fontenc}
\usepackage{ae,aecompl}
\usepackage{appendix}
\usepackage[caption=false]{subfig}
\begin{document}
\newcommand{\hi}{\mbox{H\,{\sc i}}}
\newcommand{\mgii}{\mbox{Mg\,{\sc ii}}}
\newcommand{\mgi}{\mbox{Mg\,{\sc i}}}
\newcommand{\feii}{\mbox{Fe\,{\sc ii}}}
\newcommand{\oi}{\mbox{O\,{\sc i}}}
\newcommand{\cii}{\mbox{C\,{\sc ii}}}
\newcommand{\ci}{\mbox{C\,{\sc i}}}
\newcommand{\sii}{\mbox{Si\,{\sc ii}}}
\newcommand{\znii}{\mbox{Zn~{\sc ii}}}
\newcommand{\caii}{\mbox{Ca\,{\sc ii}}}
\newcommand{\nai}{\mbox{Na\,{\sc i}}}
\def\h2{$\rm H_2$}
\def\Nh2{$N$(H${_2}$)}
\def\chin{$\chi^2_{\nu}$}
\def\chiu{$\chi_{\rm UV}$}
\def\sys{J0441$-$4313~}
\def\lya{\ensuremath{{\rm Ly}\alpha}}
\def\lymana{\ensuremath{{\rm Lyman}-\alpha}}
\def\kms{km\,s$^{-1}$}
\def\cms{cm$^{-2}$}
\def\cc{cm$^{-3}$}
\def\zabs{$z_{\rm abs}$}
\def\zem{$z_{\rm em}$}
\def\nhi{$N$($\hi$)}
\def\ln{log~$N$}
\def\nh{$n_{\rm H}$}
\def\ne{$n_{e}$}
\def\21{21-cm}
\def\ts{T$_{s}$}
\def\th{T$_{01}$}
\def\t0{T$_{0}$}
\def\ll{$\lambda\lambda$}
\def\l{$\lambda$}
\def\fc{$C_{f}$}
\def\c21{$C_{21}$}
\def\mjb{mJy\,beam$^{-1}$}
\def\taudv{$\int\tau dv$}
\def\taup{$\tau_{\rm p}$}
\def\ha{H\,$\alpha$}
\def\hb{H\,$\beta$}
\def\oi{[O\,{\sc i}]}
\def\oii{[O\,{\sc ii}]}
\def\oiii{[O\,{\sc iii}]}
\def\nii{[N\,{\sc ii}]}
\def\sii{[S\,{\sc ii}]}
\def\taudvl{$\int\tau dv^{3\sigma}_{10}$}
\def\taudv{$\int\tau dv$}
\def\vshift{$v_{\rm shift}$}
\def\wmg{$W_{\mgii}$}
\def\wfe{$W_{\feii}$}
\def\dgi{$\Delta (g-i)$}
\def\ebv{$E(B-V)$}
\def\lir{$L_{\rm IR}$}
%
%
\title[Neutral gas in merging galaxies$-$II]{
{Prevalence of neutral gas in centres of merging galaxies$-$II: nuclear \hi\ and multi-wavelength properties}
\author[R. Dutta et al.]{R. Dutta$^1$ \thanks{E-mail: rdutta@eso.org}, R. Srianand$^2$ and N. Gupta$^2$ \\ 
$^1$ European Southern Observatory, Karl-Schwarzschild-Str. 2, D-85748 Garching Near Munich, Germany \\
$^2$ Inter-University Centre for Astronomy and Astrophysics, Post Bag 4, Ganeshkhind, Pune 411007, India \\}}
\date{Accepted. Received; in original form}
\pubyear{}
\maketitle
\label{firstpage}
\pagerange{\pageref{firstpage}--\pageref{lastpage}}
%
%
\begin {abstract}  
\par\noindent
Using a sample of 38 radio-loud galaxy mergers at $z\le0.2$, we confirm the high detection rate ($\sim$84\%) of \hi\ \21\ absorption 
in mergers, which is significantly higher ($\sim$4 times) than in non-mergers. The distributions of the \hi\ column density [\nhi] 
and velocity shift of the absorption with respect to the systemic redshift of the galaxy hosting the radio source in mergers are 
significantly different from that in non-mergers. We investigate the connection of the nuclear \hi\ gas with various multi-wavelength 
properties of the mergers. While the inferred \nhi\ and gas kinematics do not show strong (i.e. $\ge3\sigma$ level) correlation with 
galaxy properties, we find that the incidence and \nhi\ of absorption tend to be slightly higher at smaller projected separations between 
the galaxy pairs and among the lower stellar mass-radio galaxies. The incidence, \nhi\ and line width of \hi\ absorption increase from 
the pre-merger to the post-merger stages. The 100\% detection rate in post-mergers indicates that the neutral gas in the circumnuclear 
regions survives the coalescence period and is not yet quenched by the nuclear radio activity.
\end {abstract}  
%
%
%
%
\section{Introduction} 
\label{sec_introduction}  
Galaxy mergers not only distort the morphology of the interacting galaxies, but as suggested by observations and predicted 
by simulations, they can have strong influence on the physical properties of the galaxies, e.g. enhancement in star formation, 
triggering of Active Galactic Nuclei (AGNs), and dilution of central metallicities \citep{kewley2006,dimatteo2007,cox2008,
ellison2008,ellison2019,scudder2012,torrey2012,satyapal2014,moreno2015,weston2017,bustamante2018}. However, untill recently, 
how the merging process impacts the physical conditions of different gas phases in the galaxies was not well-explored. In a 
recent study, \citet{ellison2018}, using \hi\ \21\ emission, have found that the atomic gas fractions in post-mergers are 
elevated compared to isolated galaxies, while \citet{pan2018} and \citet{violino2018} have found evidence for enhanced molecular 
gas fractions in galaxy pairs using CO emission. On the other hand, recent high-resolution (parsec-scale) simulation of the 
multi-phase interstellar medium in galaxy pairs shows that interactions elevate the cold-dense ($T<$300\,K; $n>$10\,\cc) gas 
mass in galaxies by $\sim$18\% \citep{moreno2019}. This gas mass remains elevated during the galaxy-pair period, i.e. for 
few Gyrs between the first and second pericentric passages. It is thus now the opportune time to investigate observationally 
the cold gas properties in different merger stages and test predictions that are becoming available from statistical studies 
using state-of-the-art simulations.

While the atomic and molecular gas has been mapped in some individual merging galaxies \citep[e.g.][]{hibbard1996,tacconi1999}, 
statistical studies of the cold gas in galaxy pairs and mergers have typically focused on the average gas properties (like gas 
mass and fraction) derived using single dish radio observations \citep[e.g.][]{ellison2015,ellison2018}. Simulations of mergers, 
however, predict concentration of stars and gas in centres of galaxies resulting from tidal torques \citep[e.g.][]{mihos1996,blumenthal2018}. 
Observations of the gas in the inner regions of galaxy mergers are, hence, very relevant to understand gas kinematics and metallicity
in the nuclear regions and fueling of nuclear activity. High spatial resolution interferometric observations of absorption against 
different radio emitting components in mergers provide an effective way of studying the cold gas within the central kilo-parsec 
regions \citep[e.g.][]{srianand2015}. The \hi\ \21\ absorption line has long been used to probe the neutral gas in radio-loud AGNs 
\citep{vangorkom1989}. In particular, it has been used to trace gas infalling and feeding the central super-massive black hole, 
as well as negative feedback in the form of outflows/winds driven by the central starbursts or AGNs \citep[e.g.][]{vermeulen2003,gupta2006,morganti2009,schulz2018}.

There have been \hi\ \21\ absorption searches in local ($z\le0.1$) luminous infrared galaxies \citep[LIRGs;][]{mirabel1988} 
and ultra-luminous infrared galaxies \citep[ULIRGs;][]{teng2013}, which are usually associated with gas-rich mergers \citep{sanders1996}. 
However, these searches were conducted with single dish radio observations. Hence, the spatial resolution was not sufficient to 
study the radio emission and \hi\ gas in the central kilo-parsec regions of these galaxies, and the absorption may be blended 
with emission in these spectra as well. These studies were also not conducted on samples selected specifically as mergers with 
strong radio-loud emission at their centres. Using radio interferometric observations of \hi\ \21\ absorption, we have recently 
conducted a pilot study of neutral gas in the central regions of a sample of galaxy mergers hosting radio-loud AGNs at $z\le0.2$ 
\citep[][hereafter D18]{dutta2018}. In this work, we expand our sample of radio-loud mergers with new \hi\ \21\ absorption measurements. 
Further, we connect the properties of the neutral gas in centres of mergers with their optical, infrared and radio properties, 
and trace the evolution of neutral gas through different merger stages. We describe the sample and observations in Section~\ref{sec_observations}. 
The results are presented and discussed in Section~\ref{sec_results}. We summarize our conclusions in Section~\ref{sec_summary}. 
We adopt a flat $\Lambda$-cold dark matter cosmology with $H_{\rm 0}$ = 70\,\kms~Mpc$^{-1}$ and $\Omega_{\rm M}$ = 0.30 throughout 
this work.
%
%
\section{Sample \& Observations}  
\label{sec_observations}  
\subsection{Sample}
\label{sec_sample}
We cross-matched visually-identified galaxy mergers \citep[see][]{darg2011,barcos2017,satyapal2017,weston2017,fu2018} 
at spectroscopic $z\le$0.2 in the Sloan Digital Sky Survey \citep[SDSS;][]{york2000} with radio sources in the Faint Images 
of the Radio Sky at Twenty-Centimeters \citep[FIRST; resolution$\sim$5$''$;][]{white1997} and the NRAO VLA Sky Survey \citep[NVSS; 
resolution$\sim$45$''$;][]{condon1998}. We thus identified 45 mergers showing radio emission with flux density greater than 
20\,mJy at 1.4\,GHz. The identification of mergers is carried out in the same way as explained in D18. We list here the selection 
criteria for visually identifying mergers in the SDSS images $-$ (i) single galaxy with disturbed central morphology and/or tails; 
(ii) single galaxy with double nuclei and/or tails; (iii) pair of interacting galaxies that show signatures of tidal disturbances.

In D18 we had studied a sub-sample of 10 mergers from the above sample, with flux density greater than 50~mJy at 1.4\,GHz. 
Here we expand our sample with \hi\ \21\ observations of 9 mergers with flux density $\sim$20-50~mJy at 1.4\,GHz. For 19 
mergers from the above sample, \hi\ \21\ observations are available in the literature (see table 5 of D18). Hence, the 
sample used here for statistical analysis comprises 38 mergers, i.e. $\sim$84\% of the above mentioned sample of $z\le$0.2 
radio-loud mergers.

From the available multi-wavelength data, we estimated various properties for the full sample (see Table~\ref{tab:fullsample}).
The stellar mass, $M_*$, is computed from SDSS photometry and the {\it kcorrect} algorithm (v\_4.2) by \citet{blanton2007} 
\citep[see for details][]{dutta2017}. The total infrared luminosity, \lir, is obtained using either Infrared Astronomical Satellite 
(IRAS) flux densities at 12, 25, 60 and 100\,$\micron$ \citep{moshir1990} following \citet{sanders1996}, or AKARI flux densities at 
90 and 140\,$\micron$ \citep{kawada2007} following \citet{takeuchi2010}. The star formation rate (SFR) is taken from the MPA-JHU SDSS 
DR7 spectroscopic catalog \citep{brinchmann2004,salim2007}. The spectral index ($\alpha_{0.15}^{1.4}$; $S_\nu \propto \nu^{\alpha}$) 
is taken from the catalog of \citet{degasperin2018}, obtained from the 147\,MHz TIFR GMRT Sky Survey (TGSS) and 1.4\,GHz NVSS fluxes. 

The median properties of the merger sample are: $z$ = 0.04, projected separation, $\rho$ = 6~kpc, radio power at 1.4\,GHz, $P_{1.4}$ = 
$3\times10^{23}$~W~Hz$^{-1}$, $\alpha_{0.15}^{1.4}$ = $-$0.48, $M_*$ = $6\times10^{10}$~M$_\odot$, \lir\ = $6\times10^{11}$~L$_\odot$, 
and SFR = 1~M$_\odot$~yr$^{-1}$. Note that we consider here only the parameters of the strong radio source whenever spatially resolved 
data of multiple objects in a merger are available. In 85\% cases, the strong radio source has higher $M_*$. For the 20 pairs where $M_*$ 
estimation is possible for both the galaxies, 80\% are major-mergers with mass ratio $\le$1:4. The mass ratios are provided in Table~\ref{tab:fullsample}. 
The galaxy pairs have typical line-of-sight velocity separation $\le500$\,\kms, with median of $\sim$100\,\kms. 72\% of the mergers are 
LIRGs (\lir\ = $10^{11-12}$~L$_\odot$) and 25\% are ULIRGs (\lir\ $\ge10^{12}$~L$_\odot$). Based on BPT diagram \citep{baldwin1981} for 
31 of the mergers with emission line measurements from SDSS spectra, 52\% of them can be classified as AGNs, 39\% as Composite and 9\% 
as star-forming. However, based on the radio power, all the sources can be classified as radio-loud AGNs. Based on SDSS colour-magnitude 
relation \citep{weinmann2006}, 58\% of the mergers are blue, 21\% are red, and 21\% are mixed (i.e. pair of blue and red galaxies).

To classify objects into different merger stages, we adapt from the six-stage merger classification scheme of the Great Observatories All-Sky 
LIRG Survey \citep{haan2011}. We combine classes 1 and 2 of this scheme as the pre-merger stage (separate galaxies with symmetric disks or 
asymmetric disks and/or tidal tails), classes 3 and 4 as the ongoing merger stage (two distinct nuclei in a common envelope or double nuclei 
with tidal tails), and classes 5 and 6 as the post-merger stage (single/obscured nucleus with long prominent tails or disturbed central 
morphology and short faint tails). There are 13, 12 and 13 mergers in the pre-merger, ongoing merger and post-merger stages, respectively.
The classification stage of each merger is also listed in Table~\ref{tab:fullsample}.
\subsection{Observations}
\label{sec_obs}
We obtained the redshifted \hi\ \21\ line observations of the nine new mergers using the Giant Metrewave Radio Telescope (GMRT) 
between October 2018 and March 2019. The observations were conducted using the L-band receivers and the GMRT Software Backend. 
A spectral set-up of 16 MHz baseband bandwidth split into 512 channels was used, which resulted in a velocity resolution 
of $\sim$7\,\kms\ and a coverage of $\sim$3600-3700\,\kms. Each source was observed for $\sim$2.5-3\,h. In addition, standard 
calibrators were regularly observed for flux density, bandpass, and gain calibrations. The data were acquired in parallel hand 
correlations. The data reduction was carried out in Astronomical Image Processing System ({\sc aips})\footnote{{\sc aips} is 
produced and maintained by the National Radio Astronomy Observatory, a facility of the National Science Foundation operated under 
cooperative agreement by Associated Universities, Inc.} following standard procedures \citep[][D18]{dutta2016}.

The GMRT 1.4\,GHz continuum images (resolution $\sim$2-3$''$) recover $\sim$80-100\% of the total flux of the radio sources 
obtained in FIRST or NVSS. The 1.4\,GHz continuum contours are overlaid on SDSS images of the mergers in Fig.~\ref{fig:overlay}. 
Four of the mergers show two radio continuum peaks. Absorption is detected from 8 out of 9 mergers. Properties of the \hi\ 
absorption spectra are listed in Table~\ref{tab:results}. The spectra along with the Gaussian fits to the absorption are shown 
in Fig.~\ref{fig:21cmspectra}. The spectra are extracted towards the radio continuum peaks. The parameters from Gaussian fits 
to the lines are given in Table~\ref{tab:gaussfit}.
%
%
\section{Results \& Discussion}
\label{sec_results}
\subsection{Incidence, \nhi\ and kinematics of \hi\ gas}
\label{sec_results1}
Our new observations reveal 8 new \hi\ \21\ absorption detections out of the 9 mergers observed. For the full sample of mergers, 
the detection rate of \hi\ \21\ absorption is $84\pm15$\% (i.e. 32/38). Note that we consider only absorption towards the strongest 
radio continuum peak associated with the merger. We obtained a reference sample of 229 non-mergers (i.e. those that do not satisfy 
the merger selection criteria listed in Section~\ref{sec_sample}) from the study of associated \hi\ absorption in $z\le0.2$ radio 
galaxies presented in \citet{maccagni2017}. The detection rate in the non-merger sample is $23\pm3$\%, i.e. four times lower than 
in the merger sample. Next, we estimate the detection rates in the merger and non-merger sample by imposing different \nhi\ or 
(\taudv) sensitivity limits (for spin temperature, \ts\ = 100~K, and covering factor, \fc\ = 1). As can be seen from Table~\ref{tab:rates}, 
the excess of incidence of \hi\ absorption in mergers compared to non-mergers increases by a factor of $\sim$4 to $\sim$38 as we 
go for higher \nhi\ sensitivity limits, implying that mergers show higher \nhi\ on average.
  
\begin{figure*}
\includegraphics[width=0.35\textwidth, angle=90]{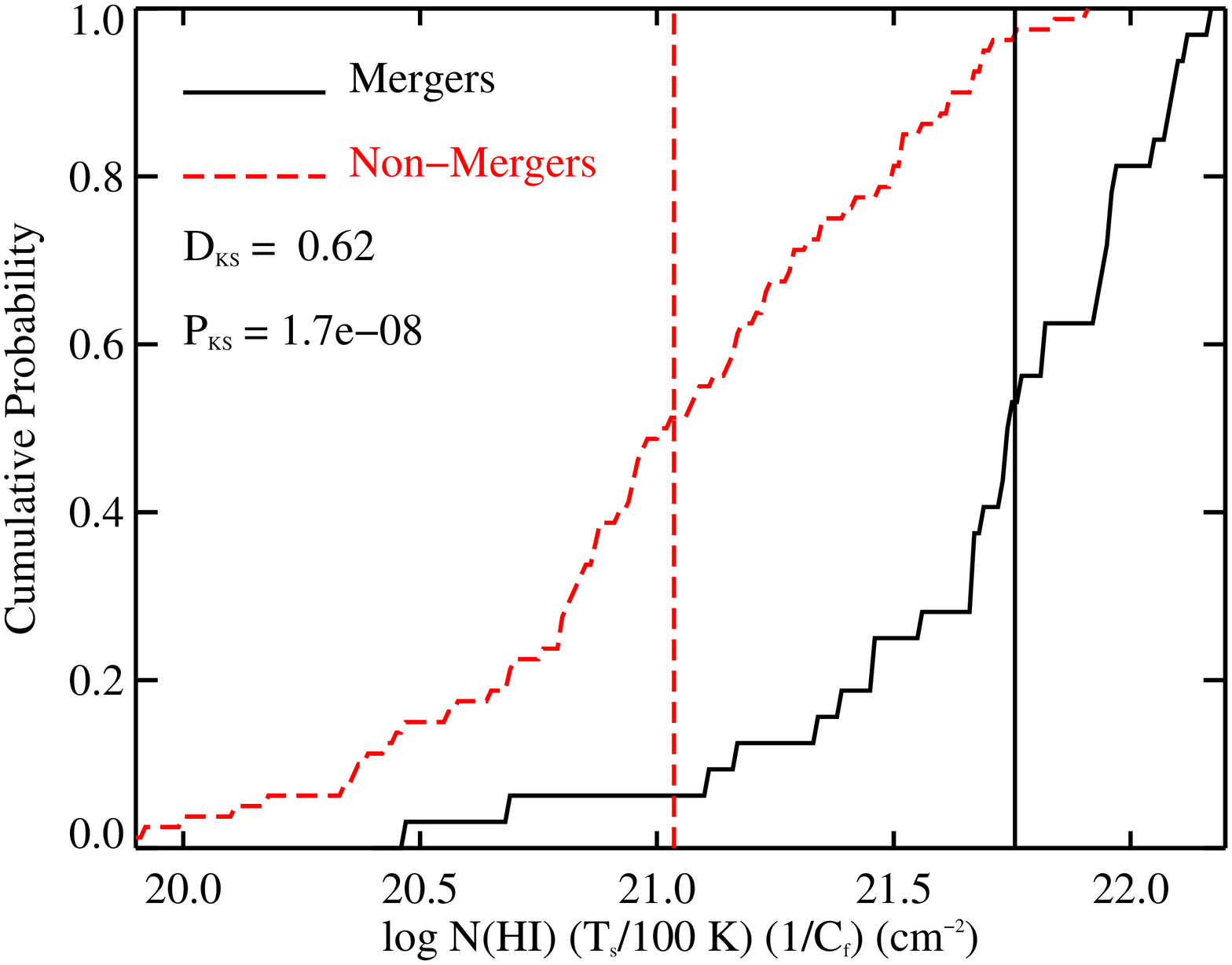}
\includegraphics[width=0.35\textwidth, angle=90]{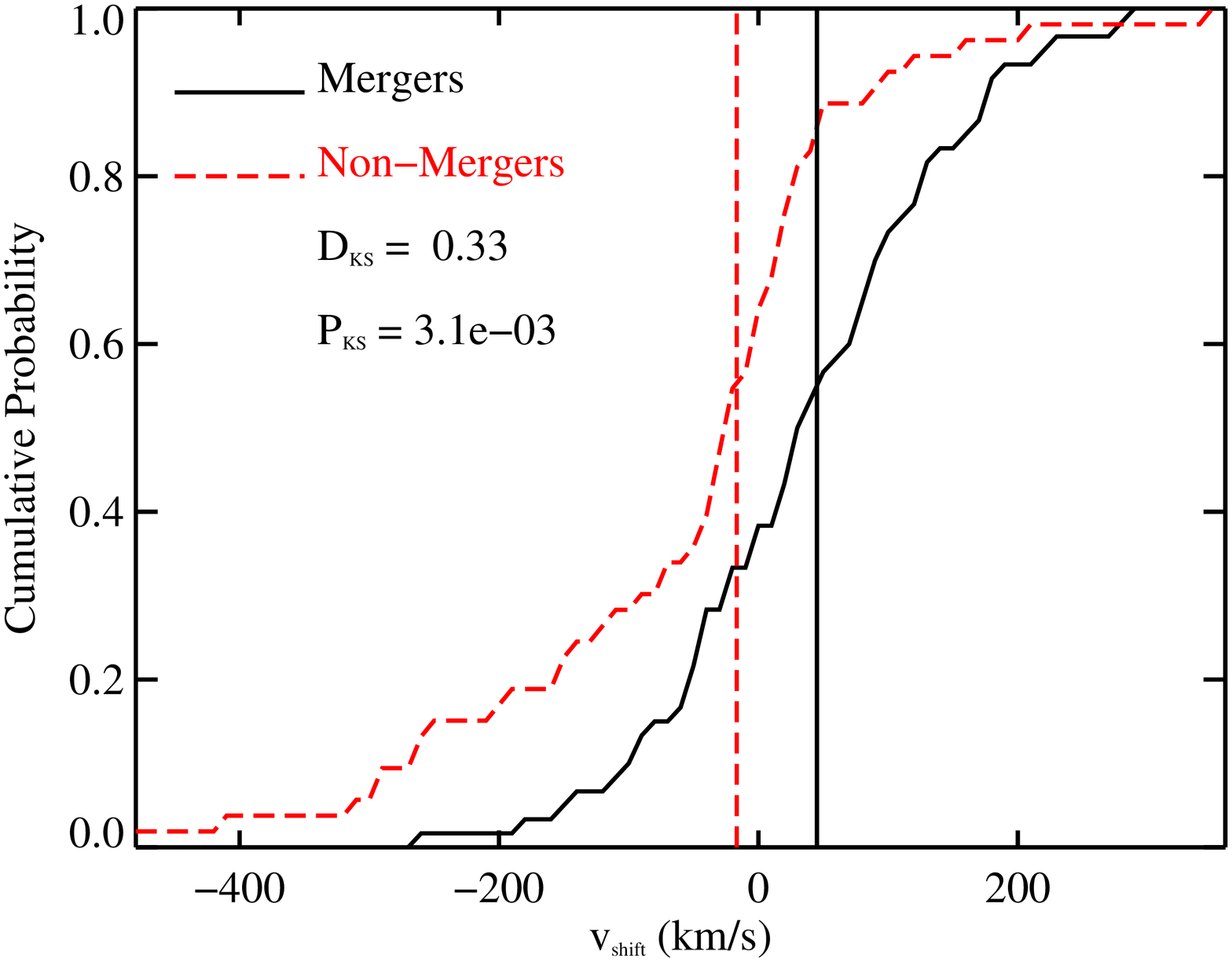}
\caption{Cumulative distributions of \nhi\ {(\it left)} and \vshift\ {(\it right)} in the sample of mergers (solid lines) and non-mergers 
(dashed lines). The median values are demarcated by solid and dashed vertical ticks for mergers and non-mergers, respectively. The distributions 
are significantly different for mergers and non-mergers, as indicated by results of two-sided KS tests.
}
\label{fig:dist}
\end{figure*}

This is confirmed by the fact that the \nhi\ distribution of mergers and non-mergers are significantly different. A two-sided 
Kolmogorov-Smirnov test shows maximum deviation between the two cumulative distributions of $D_{\rm KS}$ = 0.6, and probability 
of finding the difference by chance is $P_{\rm KS}$ = $2\times10^{-8}$. Mergers give rise to stronger absorption on average and 
have median \nhi\ five times higher than non-mergers (see Fig.~\ref{fig:dist}). In addition, we find that the distribution of 
the velocity shift (\vshift) between the \hi\ absorption components and the systemic redshift of the galaxies hosting the 
radio sources is different from that in non-mergers ($D_{\rm KS}$ = 0.3, $P_{\rm KS}$ = $3\times10^{-3}$). Mergers show three 
times more redshifted (i.e. infalling) absorption than non-mergers (Fig.~\ref{fig:dist}). The fraction of components with \vshift\ 
$\ge100$\,\kms\ is $30\pm7$\% in mergers compared to $9\pm4$\% in non-mergers. We do not find any significant difference in the 
velocity widths (full width at half optical depth; FWHM) of the absorption in mergers and non-mergers ($D_{\rm KS}$ = 0.2, 
$P_{\rm KS}$ = 0.3). Results from the comparison of \hi\ absorption properties between the samples of mergers and non-mergers 
are given in Table~\ref{tab:kstest}.

Finally, to confirm the above differences, we create a mass- and redshift-matched reference sample of non-mergers from \citet{maccagni2017}. 
In a sub-sample of 24 mergers, for each merger we randomly select a source from the non-merger sample with \hi\ \21\ detection, that is within 
$\pm$0.5~dex in $M_*$ and $\pm$0.1 in redshift, and repeat this 100 times. The resulting matched samples also have similar distributions of 
$P_{1.4}$ (median $D_{\rm KS}$ = 0.3, $P_{\rm KS}$ = 0.1). We find that the \nhi\ and \vshift\ distributions in the matched samples are 
significantly different as well (see Table~\ref{tab:kstest}). This indicates that the differences in \hi\ properties of the merger 
sample from the non-merger sample are not driven by differences in stellar mass, but rather are likely to be driven by the merging process 
feeding the central regions with large quantities of neutral hydrogen gas.
\begin{table}
\caption{Incidence of \hi\ absorption in mergers and non-mergers for different \nhi\ sensitivity limits.}
\centering
\begin{tabular}{ccc}
\hline
\nhi\ sensitivity & \multicolumn{2}{c}{Incidence of \hi\ absorption} \\
(\ts$/100$ K)($1/$\fc)(\cms) & Mergers & Non-mergers \\
\hline 
All              & 84 $\pm$ 15\% & 23 $\pm$ 3\% \\
$2\times10^{20}$ & 94 $\pm$ 17\% & 26 $\pm$ 6\% \\
$5\times10^{20}$ & 89 $\pm$ 16\% & 18 $\pm$ 3\% \\
$5\times10^{21}$ & 76 $\pm$ 17\% &  2 $\pm$ 1\% \\
\hline
\end{tabular}
\label{tab:rates}
\begin{flushleft}
\end{flushleft}
\end{table}
\begin{table}
\caption{Comparison of \hi\ absorption properties between the samples of mergers and non-mergers.}
\centering
\begin{tabular}{cccccc}
\hline
\hi\ property & \multicolumn{2}{c}{All} & \multicolumn{2}{c}{Mass-matched} \\
              & $D_{\rm KS}$ & $P_{\rm KS}$ & $D_{\rm KS}$ & $P_{\rm KS}$  \\
 (1)          & (2)          & (3)          & (4)          & (5)           \\
\hline
\nhi\    & 0.62 & $2\times10^{-8}$ & 0.70 & $3\times10^{-5}$ \\
FWHM     & 0.19 & 0.32             & 0.32 & 0.13             \\
\vshift\ & 0.33 & $3\times10^{-3}$ & 0.50 & $2\times10^{-3}$ \\
\hline
\end{tabular}
\label{tab:kstest}
\begin{flushleft}
{\it Notes.}
Column 1: \hi\ absorption property whose distributions in mergers and non-mergers are compared in a two-sided Kolmogorov-Smirnov (KS) test.
Columns 2 and 4: maximum deviation between the two cumulative distribution functions.
Columns 3 and 5: probability of finding the difference by chance. 
\end{flushleft}
\end{table}
\subsection{Implication of non-detections}
\label{sec_results2}
The only non-detection in the new sample presented here is towards the complex merger J0904$+$1435 consisting of three galaxies 
(see Fig.~\ref{fig:overlay}). The radio emission is resolved in the GMRT map, with the strongest continuum peak having $\sim$14\% 
of the total flux and not coinciding with any optical nuclei. Lack of absorption in this case could be related to the radio sightline 
not probing the nuclear region and absence of \hi\ gas in the region probed between the merging galaxies. Among the mergers presented 
here and in D18, there are four cases (J0915$+$4419, J1100$+$1002, J1315$+$6207 and J1518$+$4244), where the optical nuclei of both 
the interacting galaxies (with $\rho\sim$6-20~kpc) are co-spatial with radio emission in GMRT maps (see Fig.~\ref{fig:overlay} and 
figure 1 in D18). Absorption is detected towards only the stronger radio emission in three of them. From $3\sigma$ upper limit towards 
the weaker radio source obtained by integrating over the velocity range of absorption detected towards the stronger source, we find 
that the \hi\ optical depth towards the stronger source is higher by $\gtrsim2-4$ times. In case of the merger J1518$+$4244, absorption 
is detected towards both the radio sources, but the stronger radio source shows two times higher integrated optical depth. Three of 
these mergers also have higher SFR and dust depletion in the regions showing stronger radio emission, indicating a possible physical 
connection between the presence of \hi\ gas, starburst and nuclear radio activity. However, we note that there is no clear connection 
between the \hi\ gas and other properties like stellar mass for these four mergers, e.g. in two cases the stronger radio source showing 
higher \hi\ optical depth has the higher stellar mass of the galaxy pair and it is vice-versa for the other two.
\subsection{\hi\ gas and merger properties}
\label{sec_results3}
We checked the dependence of the incidence of \hi\ absorption on various properties of the mergers as listed in Table~\ref{tab:fullsample}. 
We find that the incidence of \hi\ is higher at smaller projected separations, i.e. $\sim$92\% at $\rho\le6$~kpc vis-\`{a}-vis $\sim$71\% 
at higher $\rho$. Detections tend to be higher in the more advanced stages of merger. Further, the incidence is higher at smaller stellar 
masses, i.e. $\sim$100\% at $M_*\le6\times10^{10}$~M$_\odot$ vis-\`{a}-vis $\sim$72\% at higher $M_*$. The non-detections in high stellar 
mass-mergers are usually associated with red elliptical galaxies. In addition, we find that the incidence is higher at lower redshifts 
($\sim$95\% at $z\le0.04$ compared to $\sim$72\% at $z>0.04$), and for flatter spectral index sources ($\sim$94\% at $\alpha_{0.15}^{1.4}>-0.48$ 
compared to $\sim$71\% at steeper $\alpha_{0.15}^{1.4}$). The dependence on spectral index could indicate higher incidence among sources 
with more compact radio emission, that is better suited to detect gas in absorption \citep{gupta2006}.

We do not find any significant dependence of incidence on radio power, infrared luminosity, SFR and nebular emission line ratios.
Majority ($\sim$97\%) of the mergers in our sample are LIRGs or ULIRGs, which have been observed to show high incidence of \hi\ 
absorption. Using single dish observations, \citet{teng2013} have found 100\% incidence of \hi\ absorption in nine ULIRGs, while
\citet{mirabel1988} have found that the incidence increases with far-infrared luminosity, from 40\% for $L_{\rm FIR}$ $\ge10^{11}$~L$_\odot$
to 100\% for $L_{\rm FIR}$ $\ge10^{12}$~L$_\odot$ in a sample of eighty galaxies. Three and eight of the mergers in our full sample are 
also present in the sample studied by \citet{teng2013} and \citet{mirabel1988}, respectively. Though as noted in Section~\ref{sec_introduction},
these samples were selected on the basis of infrared luminosity, without the constraints on morphology and radio continuum flux as in our 
merger sample. We do not find a significant difference in the incidence between LIRGs ($\sim$83\%) and ULIRGs ($\sim$88\%) in our sample. 
Though we note that our sample has three times less ULIRGs compared to LIRGs, and a larger sample is required to check if the incidence of 
nuclear \hi\ gas increases with the infrared luminosity.

Next, we carry out correlation analysis between properties of the \hi\ gas (\nhi, FWHM and \vshift) and those of the mergers using 
non-parametric Kendall's $\tau$ test (see Table~\ref{tab:correl}). We do not find any significant correlation (i.e. $\ge3\sigma$) of 
the properties of \hi\ with those of the mergers. This once again confirms that the merger-induced gas accretion is the main driver 
behind the high detection rate of \hi\ \21\ absorption. However, the fact that there is no strong relation between the galaxy 
and absorption properties tells us that it is not straightforward to connect the central gas accumulation we infer based on \hi\ \21\ 
absorption with the triggering or quenching of nuclear star formation.

\begin{figure*}
\includegraphics[width=0.25\textwidth, angle=90]{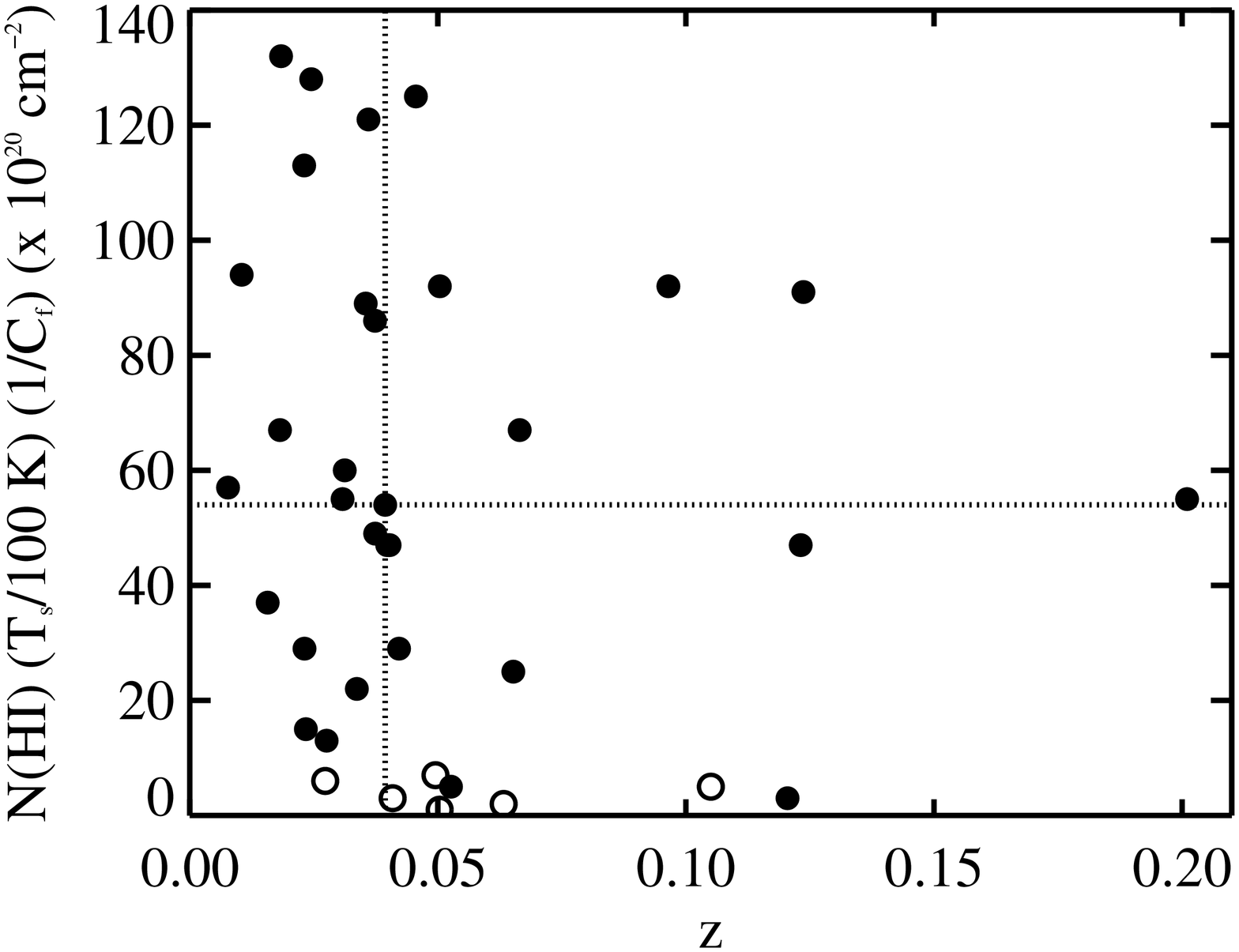}
\includegraphics[width=0.25\textwidth, angle=90]{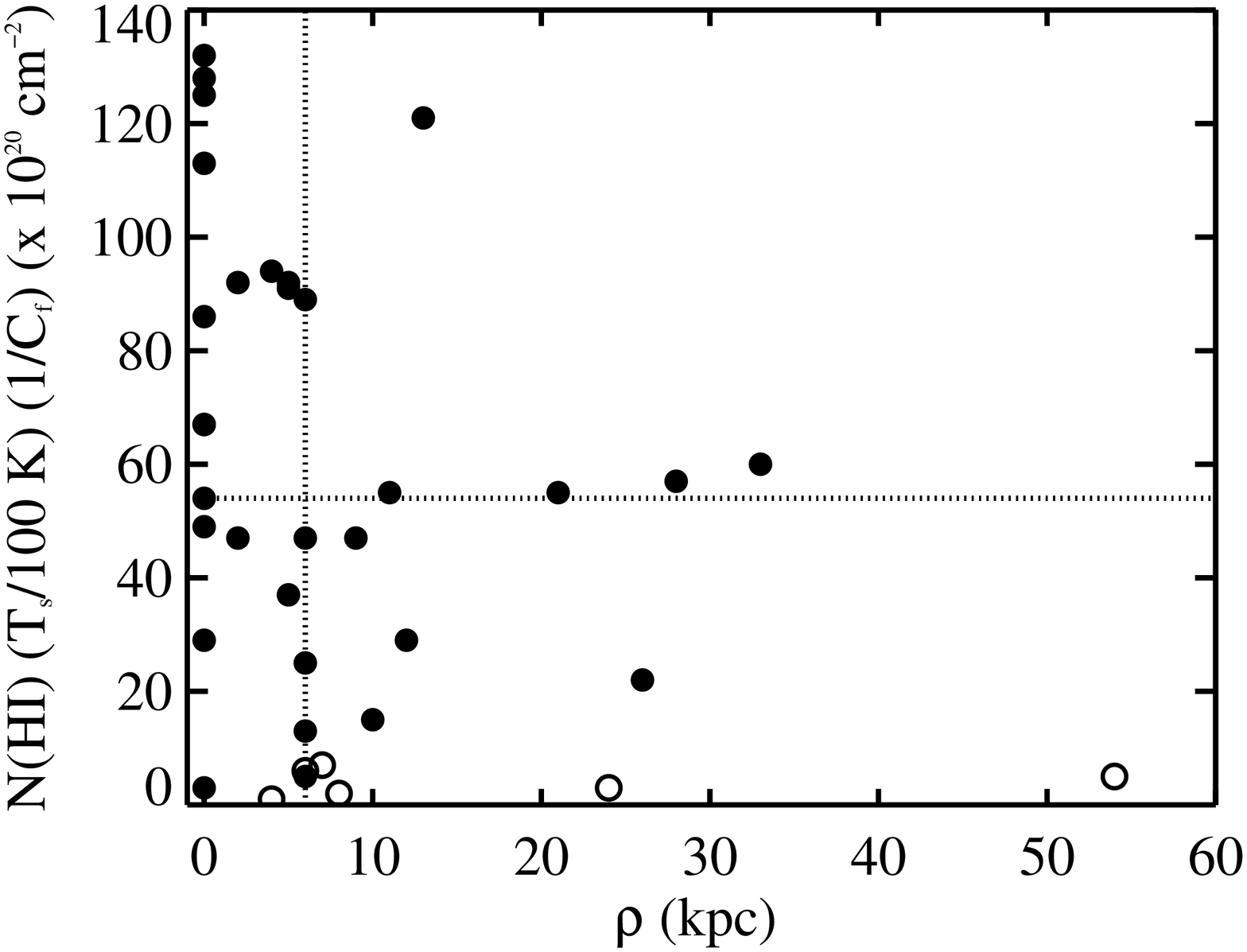}
\includegraphics[width=0.25\textwidth, angle=90]{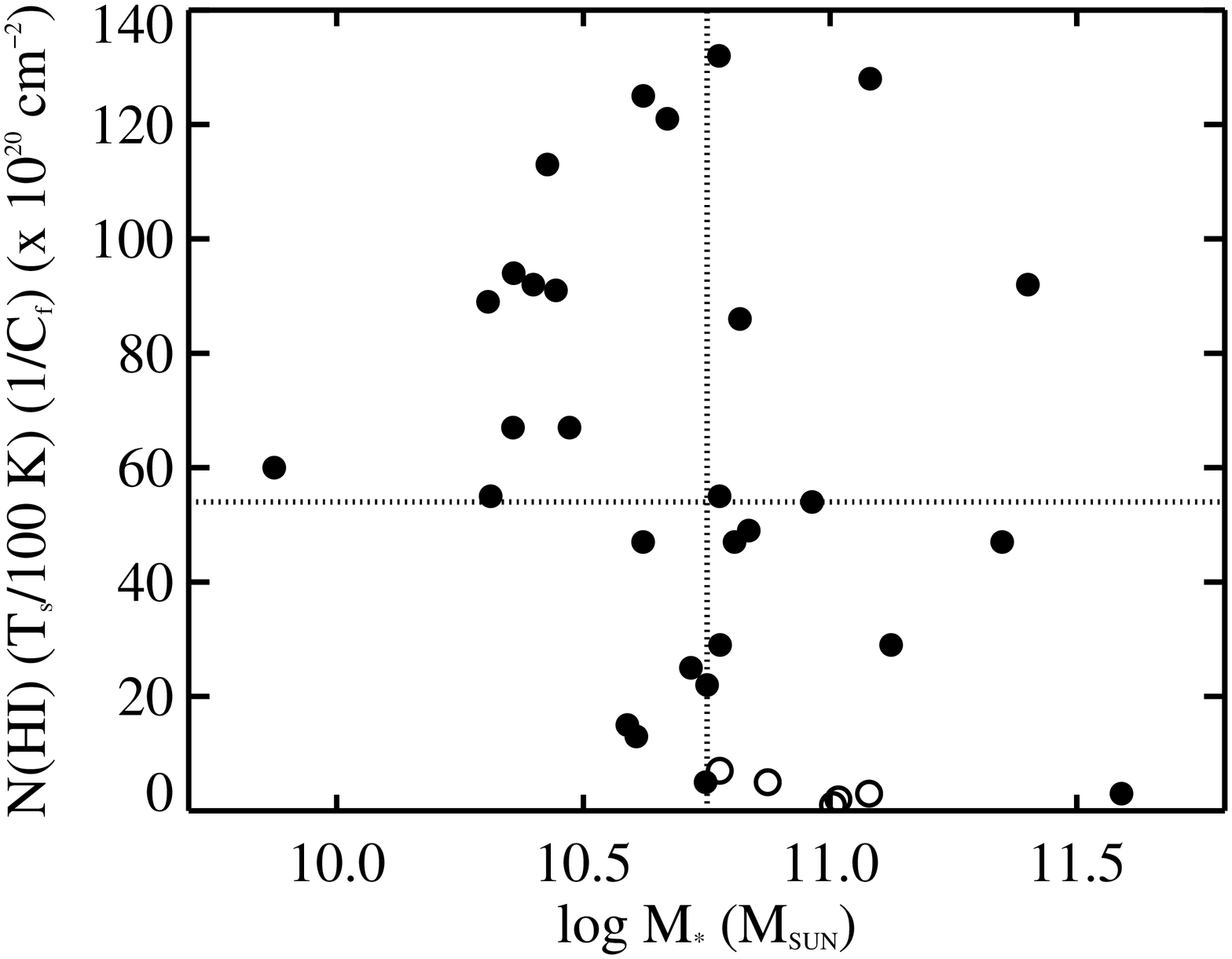}
\caption{The \nhi\ inferred for mergers as a function of redshift, projected separation and stellar mass, from left to right respectively. 
Solid symbols correspond to detections while open symbols correspond to $3\sigma$ upper limits. Median values are marked by dotted lines.
\nhi\ shows $\sim2\sigma$ anti-correlation with these parameters (see Table~\ref{tab:correl}).
}
\label{fig:correl}
\end{figure*}

We show in Fig.~\ref{fig:correl}, \nhi\ measured for the mergers as a function of $z$, $M_*$ and $\rho$, i.e. the parameters 
with which \nhi\ shows tentative ($\sim2\sigma$) anti-correlation. Increase in \nhi\ with decreasing redshift could indicate the 
presence of more gas-rich mergers at lower redshifts. However, we note that the sample lacks low stellar mass systems at high
redshifts, likely due to selection bias. Hence, the dependence of \nhi\ on redshift could be driven by its dependence on stellar 
mass. We have checked that except redshift and stellar mass, no other galaxy properties are correlated with each other.

\begin{figure}
\includegraphics[width=0.35\textwidth, angle=90]{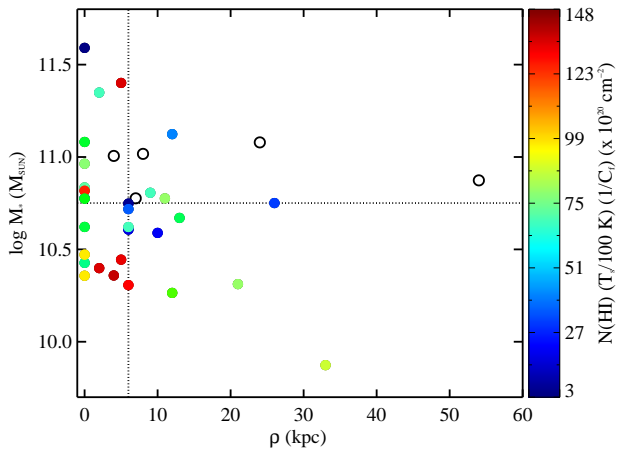}
\caption{Stellar mass versus projected separation of the mergers. Filled and open symbols represent detections (colour-coded in \nhi\ as shown in the bar to the right) 
and non-detections, respectively. The median $M_*$ and $\rho$ are marked by horizontal and vertical dashed lines, respectively. The detection rate and \nhi\ are higher 
for smaller values of $M_*$ and $\rho$.
}
\label{fig:ms_rho}
\end{figure}

Here we focus on the tentative dependences of \nhi\ on stellar mass and projected separation. Exploring the parameter space of merger properties 
considered here, we find that majority of the large \nhi\ absorption occur at smaller values of $M_*$ and $\rho$ (see Fig.~\ref{fig:ms_rho}).
We note that in the non-merger reference sample considered here, we do not find any trend of \nhi\ or detection rate with $M_*$. Therefore,
the trend of higher incidence and higher \nhi\ at lower $M_*$ in the merger sample could imply a physical connection between the merger type 
and the \hi\ cross-section. Majority ($\sim$71\%) of the lower mass mergers (i.e. below the median $M_*$) are blue in colour, while only 
$\sim$40\% of the higher mass mergers are blue. Hence, the dependence of \nhi\ on $M_*$ could be due to higher concentration of \hi\ gas in 
centres of wet or gas-rich mergers. However, this may also reflect the general trend of increasing detection rate of \hi\ emission and atomic 
gas fraction with decreasing stellar mass observed in both isolated and post-merger galaxies \citep{catinella2018,ellison2018}.

\begin{figure*}
\includegraphics[width=1.0\textwidth]{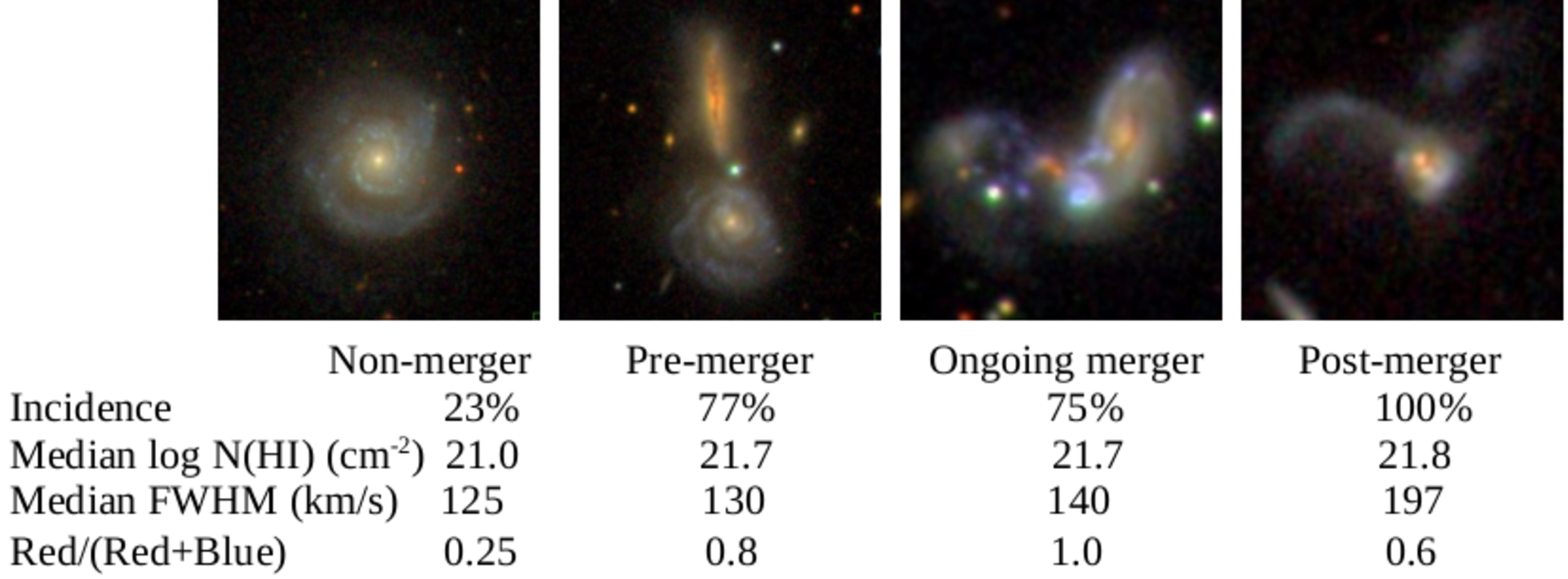}
\caption{SDSS colour composite image of a typical system in the non-merger, pre-merger, ongoing and post-merger stage, from left to right.
Below each example are listed, from top to bottom, the incidence of \hi\ absorption in that stage, and median \nhi, median FWHM and fraction
of redshifted components among all redshifted and blueshifted components for detections in that stage. 
}
\label{fig:merger_stage}
\end{figure*}

We notice that most ($\sim$94\%) of the mergers with $\rho\le6$~kpc (that show higher detection rate) are in the ongoing or post-merger stages. 
This motivated us to look at the evolution of \hi\ properties along the merger sequence. The properties of \hi\ gas in different merger stages 
are listed in Fig.~\ref{fig:merger_stage}, which shows a typical example of a merger from our sample in each of the stages. The incidence, \nhi\ 
and FWHM of \hi\ absorption increases from non-merger through post-merger stages, with 100\% detection in post-mergers. The fraction Red$/$(Red$+$Blue),
defined as the number of redshifted components (\vshift\ $\ge$100\,\kms) divided by the total number of redshifted and blueshifted (\vshift\ $\le-$100\,\kms)
components, is highest among ongoing mergers. All these are consistent with more gas flow to the central regions during different merger stages.
While the contrast between the non-merger and merger samples are statistically significant (as discussed in Section~\ref{sec_results1}), the 
differences among the three merger stages themselves are not statistically significant based on two-sided KS-tests. However, these results do 
point towards widespread presence of \hi\ gas with infall signatures throughout the merger process. 

It is particularly interesting to note that large amount of cold \hi\ gas survives in centres of all post-mergers. This could result 
out of combination of the circumnuclear gas in the progenitor galaxies as well as the merger channeling gas to the central regions. 
This leads to the conclusion that the nuclear radio activity has not yet quenched or driven away the neutral gas in the circumnuclear 
regions in the recently coalesced or post-merger stage \citep[i.e. within a $\sim$Gyr of coalescence, e.g.][]{lotz2008}. It is 
possible that depletion of neutral gas due to either nuclear winds/radio jets as predicted by simulations \citep{dimatteo2005} 
or condensation into molecular phase and consequently stars is yet to take place. On the other hand, nuclear accretion and/or 
cooling of ionized gas into neutral gas are possibly still ongoing in the centres of these post-mergers. 

The 100\% detection rate of \hi\ absorption and high values of \nhi\ that we find in post-mergers is consistent with the results 
of \citet{ellison2018}. They find higher (factor of $\sim$1.5) detection rate of \hi\ emission as well as atomic gas fraction (factor 
of $\sim$3) in post-mergers compared to a control sample of isolated galaxies matched in stellar mass. Thus the merger process leads
to increase in both the concentration of nuclear \hi\ gas as well as the overall neutral gas fraction in post-mergers. Based on a revised 
picture of the merger process, \citet{ellison2018} rule out a `blowout' phase in the merger-driven sequence of galaxy evolution \citep{hopkins2008}. 
Our results support this and indicate that the circumnuclear \hi\ gas probably survives till the `quasar' phase in this evolutionary 
picture. This is further corroborated by the results of \citet{teng2013}, who find 100\% incidence of \hi\ absorption in coalesced 
ULIRGs compared to $\sim$60\% in far-infrared-weak quasars. In addition, while only $\sim$25\% of the post-mergers in \citet{ellison2018}
are classified as optical AGNs, all our post-mergers host radio-loud AGNs at their centres. This implies that the AGN activity in the 
post-merger stage has not yet consumed or expelled the \hi\ gas fed to the nuclear regions by the merger, which could also have triggered
it in the first place.
\begin{table*} 
\caption{Correlation analysis between \hi\ absorption and properties of the mergers.}
\centering
\begin{tabular}{ccccccccccccccc}
\hline
Parameter & \multicolumn{5}{c}{\nhi}            & \multicolumn{5}{c}{FWHM}            & \multicolumn{4}{c}{\vshift}        \\
          & $N$ & $r_k$ & $P(r_k)$ & $S(r_k)$ & & $N$ & $r_k$ & $P(r_k)$ & $S(r_k)$ & & $N$  & $r_k$ & $P(r_k)$ & $S(r_k)$ \\
(1)       & (2) & (3)   & (4)      & (5)      & & (6) & (7)   & (8)      & (9)      & & (10) & (11)  & (12)     & (13)     \\
\hline    
$z$                   & 38 & {\bf$-$0.19} & {\bf0.10} & {\bf1.7$\sigma$} & ~~ & 32 & $-$0.16      & 0.20       & 1.3$\sigma$      & ~~ & 32 & $-$0.03      & 0.82      & 0.2$\sigma$      \\
$\rho$                & 38 & {\bf$-$0.21} & {\bf0.06} & {\bf1.9$\sigma$} & ~~ & 32 & {\bf$-$0.20} & {\bf0.11}  & {\bf1.6$\sigma$} & ~~ & 32 & 0.16         & 0.20      & 1.3$\sigma$      \\
$P_{1.4}$             & 38 & $-$0.11      & 0.36      & 0.9$\sigma$      & ~~ & 32 & $-$0.03      & 0.83       & 0.2$\sigma$      & ~~ & 32 & $-$0.13      & 0.30      & 1.0$\sigma$      \\
$\alpha_{0.15}^{1.4}$ & 33 & 0.04         & 0.76      & 0.3$\sigma$      & ~~ & 27 & 0.05         & 0.72       & 0.4$\sigma$      & ~~ & 27 & {\bf$-$0.26} & {\bf0.05} & {\bf1.9$\sigma$} \\
$M_*$                 & 35 & {\bf$-$0.28} & {\bf0.02} & {\bf2.4$\sigma$} & ~~ & 30 & 0.02         & 0.86       & 0.2$\sigma$      & ~~ & 30 & $-$0.15      & 0.26      & 1.1$\sigma$      \\
\lir\                 & 32 & 0.01         & 1.00      & 0.1$\sigma$      & ~~ & 27 & {\bf0.26}    & {\bf0.05}  & {\bf1.9$\sigma$} & ~~ & 27 & $-$0.06      & 0.68      & 0.4$\sigma$      \\
SFR                   & 31 & $-$0.12      & 0.37      & 0.9$\sigma$      & ~~ & 26 & 0.07         & 0.60       & 0.5$\sigma$      & ~~ & 26 & {\bf0.26}    & {\bf0.06} & {\bf1.9$\sigma$} \\
\hline
\end{tabular}
\label{tab:correl}
\begin{flushleft}
Column 1: property of merger with which correlation of \nhi, FWHM and \vshift\ of the absorption is tested. 
Columns 2, 6, 10: number of measurements included in test. 
Columns 3, 7, 11: Kendall rank correlation coefficient. 
Columns 4, 8, 12: probability of correlation arising by chance. 
Columns 5, 9, 13: significance of correlation assuming Gaussian statistics. 
Upper limits of \nhi\ are considered as censored data points during survival analysis with {\tt cenken} function in {\tt NADA} package in {\tt R}.
Tentative correlations are marked in bold.
\end{flushleft}
\end{table*}
%
%
\section{Conclusion}  
\label{sec_summary} 
We have presented a study of \hi\ \21\ absorbing gas in central regions of 38 $z\le0.2$ radio-loud galaxy mergers. 
We confirm that the merger sample, in comparison to a reference sample of non-merging radio galaxies, show higher: 
(i) incidence of \hi\ \21\ absorption by a factor of $\sim$4, (ii) \nhi\ by a factor of $\sim$5, and (iii) infall signature,
in the form of redshifted absorption with respect to the systemic redshift, by a factor of $\sim$3. These differences 
persist in the redshift- and stellar mass-matched samples of mergers and non-mergers as well.

We further analyze the dependence of the incidence, \nhi\ and kinematics of the nuclear \hi\ gas in mergers on different
optical, infrared and radio properties of the mergers, as well as the evolution of the nuclear \hi\ gas along the merger 
sequence. We do not find any significant ($\ge3\sigma$) correlation of the \hi\ absorption properties with the global 
properties of the mergers. From comparison with stellar mass-matched reference sample of non-mergers and lack of strong 
dependence of \hi\ gas on any properties of the galaxies, we conclude that the merging process is likely to be the dominant 
factor behind the prevalence of high \nhi\ absorption with signature of infall among mergers. We do find tentative trend 
of increasing incidence and \nhi\ with decreasing stellar mass and projected separation. This could be due to presence of
more \hi\ gas in central regions of blue gas-rich mergers and in later stages of mergers. 

We find increasing trend of incidence, \nhi\ and velocity width from the pre-merger to the post-merger stages, though the 
differences in \hi\ gas properties between the merger stages are not statistically significant. The results presented here 
could imply evolution in the \hi\ gas properties as the merger progresses, which is relevant to explore further with larger 
samples of mergers. Higher spatial resolution resolved spectroscopy is also crucial to establish the connection of the \hi\ 
gas with nuclear starburst and radio activity. \\ 

%
%
\noindent \textbf{ACKNOWLEDGEMENTS} \newline \newline
\noindent 
We thank the anonymous reviewer for their constructive comments.
RD acknowledges support from the Alexander von Humboldt Foundation.
We thank the staff at GMRT for their help during the observations. 
GMRT is run by the National Centre for Radio Astrophysics of the Tata Institute of Fundamental Research. 

Funding for SDSS-III has been provided by the Alfred P. Sloan Foundation, the Participating Institutions, the National Science Foundation, 
and the U.S. Department of Energy Office of Science. The SDSS-III web site is http://www.sdss3.org/.
SDSS-III is managed by the Astrophysical Research Consortium for the Participating Institutions of the SDSS-III Collaboration 
including the University of Arizona, the Brazilian Participation Group, Brookhaven National Laboratory, Carnegie Mellon University, 
University of Florida, the French Participation Group, the German Participation Group, Harvard University, the Instituto de Astrofisica 
de Canarias, the Michigan State/Notre Dame/JINA Participation Group, Johns Hopkins University, Lawrence Berkeley National Laboratory, 
Max Planck Institute for Astrophysics, Max Planck Institute for Extraterrestrial Physics, New Mexico State University, New York University, 
Ohio State University, Pennsylvania State University, University of Portsmouth, Princeton University, the Spanish Participation Group, 
University of Tokyo, University of Utah, Vanderbilt University, University of Virginia, University of Washington, and Yale University.
%
%
\def\aj{AJ}%
\def\actaa{Acta Astron.}%
\def\araa{ARA\&A}%
\def\apj{ApJ}%
\def\apjl{ApJ}%
\def\apjs{ApJS}%
\def\ao{Appl.~Opt.}%
\def\apss{Ap\&SS}%
\def\aap{A\&A}%
\def\aapr{A\&A~Rev.}%
\def\aaps{A\&AS}%
\def\azh{A$Z$h}%
\def\baas{BAAS}%
\def\bac{Bull. astr. Inst. Czechosl.}%
\def\caa{Chinese Astron. Astrophys.}%
\def\cjaa{Chinese J. Astron. Astrophys.}%
\def\icarus{Icarus}%
\def\jcap{J. Cosmology Astropart. Phys.}%
\def\jrasc{JRASC}%
\def\mnras{MNRAS}%
\def\memras{MmRAS}%
\def\na{New A}%
\def\nar{New A Rev.}%
\def\pasa{PASA}%
\def\pra{Phys.~Rev.~A}%
\def\prb{Phys.~Rev.~B}%
\def\prc{Phys.~Rev.~C}%
\def\prd{Phys.~Rev.~D}%
\def\pre{Phys.~Rev.~E}%
\def\prl{Phys.~Rev.~Lett.}%
\def\pasp{PASP}%
\def\pasj{PASJ}%
\def\qjras{QJRAS}%
\def\rmxaa{Rev. Mexicana Astron. Astrofis.}%
\def\skytel{S\&T}%
\def\solphys{Sol.~Phys.}%
\def\sovast{Soviet~Ast.}%
\def\ssr{Space~Sci.~Rev.}%
\def\zap{$Z$Ap}%
\def\nat{Nature}%
\def\iaucirc{IAU~Circ.}%
\def\aplett{Astrophys.~Lett.}%
\def\apspr{Astrophys.~Space~Phys.~Res.}%
\def\bain{Bull.~Astron.~Inst.~Netherlands}%
\def\fcp{Fund.~Cosmic~Phys.}%
\def\gca{Geochim.~Cosmochim.~Acta}%
\def\grl{Geophys.~Res.~Lett.}%
\def\jcp{J.~Chem.~Phys.}%
\def\jgr{J.~Geophys.~Res.}%
\def\jqsrt{J.~Quant.~Spec.~Radiat.~Transf.}%
\def\memsai{Mem.~Soc.~Astron.~Italiana}%
\def\nphysa{Nucl.~Phys.~A}%
\def\physrep{Phys.~Rep.}%
\def\physscr{Phys.~Scr}%
\def\planss{Planet.~Space~Sci.}%
\def\procspie{Proc.~SPIE}%
\let\astap=\aap
\let\apjlett=\apjl
\let\apjsupp=\apjs
\let\applopt=\ao
\bibliographystyle{mnras}
\bibliography{mybib}

\begin{thebibliography}{}
\makeatletter
\relax
\def\mn@urlcharsother{\let\do\@makeother \do\$\do\&\do\#\do\^\do\_\do\%\do\~}
\def\mn@doi{\begingroup\mn@urlcharsother \@ifnextchar [ {\mn@doi@}
  {\mn@doi@[]}}
\def\mn@doi@[#1]#2{\def\@tempa{#1}\ifx\@tempa\@empty \href
  {http://dx.doi.org/#2} {doi:#2}\else \href {http://dx.doi.org/#2} {#1}\fi
  \endgroup}
\def\mn@eprint#1#2{\mn@eprint@#1:#2::\@nil}
\def\mn@eprint@arXiv#1{\href {http://arxiv.org/abs/#1} {{\tt arXiv:#1}}}
\def\mn@eprint@dblp#1{\href {http://dblp.uni-trier.de/rec/bibtex/#1.xml}
  {dblp:#1}}
\def\mn@eprint@#1:#2:#3:#4\@nil{\def\@tempa {#1}\def\@tempb {#2}\def\@tempc
  {#3}\ifx \@tempc \@empty \let \@tempc \@tempb \let \@tempb \@tempa \fi \ifx
  \@tempb \@empty \def\@tempb {arXiv}\fi \@ifundefined
  {mn@eprint@\@tempb}{\@tempb:\@tempc}{\expandafter \expandafter \csname
  mn@eprint@\@tempb\endcsname \expandafter{\@tempc}}}

\bibitem[\protect\citeauthoryear{{Baan}, {Hagiwara}  \& {Hofner}}{{Baan}
  et~al.}{2007}]{baan2007}
{Baan} W.~A.,  {Hagiwara} Y.,   {Hofner} P.,  2007, \mn@doi [\apj]
  {10.1086/513593}, \href {http://adsabs.harvard.edu/abs/2007ApJ...661..173B}
  {661, 173}

\bibitem[\protect\citeauthoryear{{Baldwin}, {Phillips}  \&
  {Terlevich}}{{Baldwin} et~al.}{1981}]{baldwin1981}
{Baldwin} J.~A.,  {Phillips} M.~M.,   {Terlevich} R.,  1981, \mn@doi [\pasp]
  {10.1086/130766}, \href {http://adsabs.harvard.edu/abs/1981PASP...93....5B}
  {93, 5}

\bibitem[\protect\citeauthoryear{{Barcos-Mu{\~n}oz} et~al.,}{{Barcos-Mu{\~n}oz}
  et~al.}{2017}]{barcos2017}
{Barcos-Mu{\~n}oz} L.,  et~al., 2017, \mn@doi [\apj]
  {10.3847/1538-4357/aa789a}, \href
  {https://ui.adsabs.harvard.edu/abs/2017ApJ...843..117B} {843, 117}

\bibitem[\protect\citeauthoryear{{Blanton} \& {Roweis}}{{Blanton} \&
  {Roweis}}{2007}]{blanton2007}
{Blanton} M.~R.,  {Roweis} S.,  2007, \mn@doi [\aj] {10.1086/510127}, \href
  {http://adsabs.harvard.edu/abs/2007AJ....133..734B} {133, 734}

\bibitem[\protect\citeauthoryear{{Blumenthal} \& {Barnes}}{{Blumenthal} \&
  {Barnes}}{2018}]{blumenthal2018}
{Blumenthal} K.~A.,  {Barnes} J.~E.,  2018, \mn@doi [\mnras]
  {10.1093/mnras/sty1605}, \href
  {https://ui.adsabs.harvard.edu/abs/2018MNRAS.479.3952B} {479, 3952}

\bibitem[\protect\citeauthoryear{{Brinchmann}, {Charlot}, {White}, {Tremonti},
  {Kauffmann}, {Heckman}  \& {Brinkmann}}{{Brinchmann}
  et~al.}{2004}]{brinchmann2004}
{Brinchmann} J.,  {Charlot} S.,  {White} S.~D.~M.,  {Tremonti} C.,  {Kauffmann}
  G.,  {Heckman} T.,   {Brinkmann} J.,  2004, \mn@doi [\mnras]
  {10.1111/j.1365-2966.2004.07881.x}, \href
  {http://cdsads.u-strasbg.fr/abs/2004MNRAS.351.1151B} {351, 1151}

\bibitem[\protect\citeauthoryear{{Bustamante}, {Sparre}, {Springel}  \&
  {Grand}}{{Bustamante} et~al.}{2018}]{bustamante2018}
{Bustamante} S.,  {Sparre} M.,  {Springel} V.,   {Grand} R.~J.~J.,  2018,
  \mn@doi [\mnras] {10.1093/mnras/sty1692}, \href
  {http://adsabs.harvard.edu/abs/2018MNRAS.479.3381B} {479, 3381}

\bibitem[\protect\citeauthoryear{{Carilli}, {Wrobel}  \& {Ulvestad}}{{Carilli}
  et~al.}{1998}]{carilli1998b}
{Carilli} C.~L.,  {Wrobel} J.~M.,   {Ulvestad} J.~S.,  1998, \mn@doi [\aj]
  {10.1086/300253}, \href {http://adsabs.harvard.edu/abs/1998AJ....115..928C}
  {115, 928}

\bibitem[\protect\citeauthoryear{{Catinella} et~al.,}{{Catinella}
  et~al.}{2018}]{catinella2018}
{Catinella} B.,  et~al., 2018, \mn@doi [\mnras] {10.1093/mnras/sty089}, \href
  {http://adsabs.harvard.edu/abs/2018MNRAS.476..875C} {476, 875}

\bibitem[\protect\citeauthoryear{{Chandola}, {Sirothia}  \&
  {Saikia}}{{Chandola} et~al.}{2011}]{chandola2011}
{Chandola} Y.,  {Sirothia} S.~K.,   {Saikia} D.~J.,  2011, \mn@doi [\mnras]
  {10.1111/j.1365-2966.2011.19607.x}, \href
  {http://adsabs.harvard.edu/abs/2011MNRAS.418.1787C} {418, 1787}

\bibitem[\protect\citeauthoryear{{Chandola}, {Sirothia}, {Saikia}  \&
  {Gupta}}{{Chandola} et~al.}{2012}]{chandola2012}
{Chandola} Y.,  {Sirothia} S.~K.,  {Saikia} D.~J.,   {Gupta} N.,  2012,
  Bulletin of the Astronomical Society of India, \href
  {http://adsabs.harvard.edu/abs/2012BASI...40..139C} {40, 139}

\bibitem[\protect\citeauthoryear{{Condon}, {Cotton}, {Greisen}, {Yin},
  {Perley}, {Taylor}  \& {Broderick}}{{Condon} et~al.}{1998}]{condon1998}
{Condon} J.~J.,  {Cotton} W.~D.,  {Greisen} E.~W.,  {Yin} Q.~F.,  {Perley}
  R.~A.,  {Taylor} G.~B.,   {Broderick} J.~J.,  1998, \mn@doi [\aj]
  {10.1086/300337}, \href {http://adsabs.harvard.edu/abs/1998AJ....115.1693C}
  {115, 1693}

\bibitem[\protect\citeauthoryear{{Cox}, {Jonsson}, {Somerville}, {Primack}  \&
  {Dekel}}{{Cox} et~al.}{2008}]{cox2008}
{Cox} T.~J.,  {Jonsson} P.,  {Somerville} R.~S.,  {Primack} J.~R.,   {Dekel}
  A.,  2008, \mn@doi [\mnras] {10.1111/j.1365-2966.2007.12730.x}, \href
  {http://adsabs.harvard.edu/abs/2008MNRAS.384..386C} {384, 386}

\bibitem[\protect\citeauthoryear{{Darg}, {Kaviraj}, {Lintott}, {Schawinski},
  {Silk}, {Lynn}, {Bamford}  \& {Nichol}}{{Darg} et~al.}{2011}]{darg2011}
{Darg} D.~W.,  {Kaviraj} S.,  {Lintott} C.~J.,  {Schawinski} K.,  {Silk} J.,
  {Lynn} S.,  {Bamford} S.,   {Nichol} R.~C.,  2011, \mn@doi [\mnras]
  {10.1111/j.1365-2966.2011.18964.x}, \href
  {https://ui.adsabs.harvard.edu/abs/2011MNRAS.416.1745D} {416, 1745}

\bibitem[\protect\citeauthoryear{{Di Matteo}, {Springel}  \& {Hernquist}}{{Di
  Matteo} et~al.}{2005}]{dimatteo2005}
{Di Matteo} T.,  {Springel} V.,   {Hernquist} L.,  2005, \mn@doi [\nat]
  {10.1038/nature03335}, \href
  {http://adsabs.harvard.edu/abs/2005Natur.433..604D} {433, 604}

\bibitem[\protect\citeauthoryear{{Di Matteo}, {Combes}, {Melchior}  \&
  {Semelin}}{{Di Matteo} et~al.}{2007}]{dimatteo2007}
{Di Matteo} P.,  {Combes} F.,  {Melchior} A.-L.,   {Semelin} B.,  2007, \mn@doi
  [\aap] {10.1051/0004-6361:20066959}, \href
  {http://adsabs.harvard.edu/abs/2007A%26A...468...61D} {468, 61}

\bibitem[\protect\citeauthoryear{{Dickey}}{{Dickey}}{1986}]{dickey1986}
{Dickey} J.~M.,  1986, \mn@doi [\apj] {10.1086/163793}, \href
  {http://adsabs.harvard.edu/abs/1986ApJ...300..190D} {300, 190}

\bibitem[\protect\citeauthoryear{{Dutta}, {Gupta}, {Srianand}  \&
  {O'Meara}}{{Dutta} et~al.}{2016}]{dutta2016}
{Dutta} R.,  {Gupta} N.,  {Srianand} R.,   {O'Meara} J.~M.,  2016, \mn@doi
  [\mnras] {10.1093/mnras/stv2980}, \href
  {http://adsabs.harvard.edu/abs/2016MNRAS.456.4209D} {456, 4209}

\bibitem[\protect\citeauthoryear{{Dutta}, {Srianand}, {Gupta}, {Momjian},
  {Noterdaeme}, {Petitjean}  \& {Rahmani}}{{Dutta} et~al.}{2017}]{dutta2017}
{Dutta} R.,  {Srianand} R.,  {Gupta} N.,  {Momjian} E.,  {Noterdaeme} P.,
  {Petitjean} P.,   {Rahmani} H.,  2017, \mn@doi [\mnras]
  {10.1093/mnras/stw2689}, \href
  {http://adsabs.harvard.edu/abs/2017MNRAS.465..588D} {465, 588}

\bibitem[\protect\citeauthoryear{{Dutta}, {Srianand}  \& {Gupta}}{{Dutta}
  et~al.}{2018}]{dutta2018}
{Dutta} R.,  {Srianand} R.,   {Gupta} N.,  2018, \mn@doi [\mnras]
  {10.1093/mnras/sty1872}, \href
  {http://adsabs.harvard.edu/abs/2018MNRAS.480..947D} {480, 947}

\bibitem[\protect\citeauthoryear{{Ellison}, {Patton}, {Simard}  \&
  {McConnachie}}{{Ellison} et~al.}{2008}]{ellison2008}
{Ellison} S.~L.,  {Patton} D.~R.,  {Simard} L.,   {McConnachie} A.~W.,  2008,
  \mn@doi [\aj] {10.1088/0004-6256/135/5/1877}, \href
  {http://adsabs.harvard.edu/abs/2008AJ....135.1877E} {135, 1877}

\bibitem[\protect\citeauthoryear{{Ellison}, {Fertig}, {Rosenberg}, {Nair},
  {Simard}, {Torrey}  \& {Patton}}{{Ellison} et~al.}{2015}]{ellison2015}
{Ellison} S.~L.,  {Fertig} D.,  {Rosenberg} J.~L.,  {Nair} P.,  {Simard} L.,
  {Torrey} P.,   {Patton} D.~R.,  2015, \mn@doi [\mnras]
  {10.1093/mnras/stu2744}, \href
  {http://adsabs.harvard.edu/abs/2015MNRAS.448..221E} {448, 221}

\bibitem[\protect\citeauthoryear{{Ellison}, {Catinella}  \&
  {Cortese}}{{Ellison} et~al.}{2018}]{ellison2018}
{Ellison} S.~L.,  {Catinella} B.,   {Cortese} L.,  2018, \mn@doi [\mnras]
  {10.1093/mnras/sty1247}, \href
  {http://adsabs.harvard.edu/abs/2018MNRAS.478.3447E} {478, 3447}

\bibitem[\protect\citeauthoryear{{Ellison}, {Viswanathan}, {Patton},
  {Bottrell}, {McConnachie}, {Gwyn}  \& {Cuillandre}}{{Ellison}
  et~al.}{2019}]{ellison2019}
{Ellison} S.~L.,  {Viswanathan} A.,  {Patton} D.~R.,  {Bottrell} C.,
  {McConnachie} A.~W.,  {Gwyn} S.,   {Cuillandre} J.-C.,  2019, \mn@doi
  [\mnras] {10.1093/mnras/stz1431}, \href
  {https://ui.adsabs.harvard.edu/abs/2019MNRAS.tmp.1374E} {p.~1374}

\bibitem[\protect\citeauthoryear{{Fu} et~al.,}{{Fu} et~al.}{2018}]{fu2018}
{Fu} H.,  et~al., 2018, \mn@doi [\apj] {10.3847/1538-4357/aab364}, \href
  {https://ui.adsabs.harvard.edu/abs/2018ApJ...856...93F} {856, 93}

\bibitem[\protect\citeauthoryear{{Gallimore}, {Baum}, {O'Dea}, {Pedlar}  \&
  {Brinks}}{{Gallimore} et~al.}{1999}]{gallimore1999}
{Gallimore} J.~F.,  {Baum} S.~A.,  {O'Dea} C.~P.,  {Pedlar} A.,   {Brinks} E.,
  1999, \mn@doi [\apj] {10.1086/307853}, \href
  {http://adsabs.harvard.edu/abs/1999ApJ...524..684G} {524, 684}

\bibitem[\protect\citeauthoryear{{Ger{\'e}b}, {Maccagni}, {Morganti}  \&
  {Oosterloo}}{{Ger{\'e}b} et~al.}{2015}]{gereb2015}
{Ger{\'e}b} K.,  {Maccagni} F.~M.,  {Morganti} R.,   {Oosterloo} T.~A.,  2015,
  \mn@doi [\aap] {10.1051/0004-6361/201424655}, \href
  {http://adsabs.harvard.edu/abs/2015A%26A...575A..44G} {575, A44}

\bibitem[\protect\citeauthoryear{{Gupta}, {Salter}, {Saikia}, {Ghosh}  \&
  {Jeyakumar}}{{Gupta} et~al.}{2006}]{gupta2006}
{Gupta} N.,  {Salter} C.~J.,  {Saikia} D.~J.,  {Ghosh} T.,   {Jeyakumar} S.,
  2006, \mn@doi [\mnras] {10.1111/j.1365-2966.2006.11064.x}, \href
  {http://adsabs.harvard.edu/abs/2006MNRAS.373..972G} {373, 972}

\bibitem[\protect\citeauthoryear{{Haan} et~al.,}{{Haan}
  et~al.}{2011}]{haan2011}
{Haan} S.,  et~al., 2011, \mn@doi [\aj] {10.1088/0004-6256/141/3/100}, \href
  {http://adsabs.harvard.edu/abs/2011AJ....141..100H} {141, 100}

\bibitem[\protect\citeauthoryear{{Hibbard} \& {van Gorkom}}{{Hibbard} \& {van
  Gorkom}}{1996}]{hibbard1996}
{Hibbard} J.~E.,  {van Gorkom} J.~H.,  1996, \mn@doi [\aj] {10.1086/117815},
  \href {http://adsabs.harvard.edu/abs/1996AJ....111..655H} {111, 655}

\bibitem[\protect\citeauthoryear{{Hopkins}, {Hernquist}, {Cox}  \& {Kere{\v
  s}}}{{Hopkins} et~al.}{2008}]{hopkins2008}
{Hopkins} P.~F.,  {Hernquist} L.,  {Cox} T.~J.,   {Kere{\v s}} D.,  2008,
  \mn@doi [\apjs] {10.1086/524362}, \href
  {http://adsabs.harvard.edu/abs/2008ApJS..175..356H} {175, 356}

\bibitem[\protect\citeauthoryear{{Kawada} et~al.,}{{Kawada}
  et~al.}{2007}]{kawada2007}
{Kawada} M.,  et~al., 2007, \mn@doi [\pasj] {10.1093/pasj/59.sp2.S389}, \href
  {http://adsabs.harvard.edu/abs/2007PASJ...59S.389K} {59, S389}

\bibitem[\protect\citeauthoryear{{Kewley}, {Geller}  \& {Barton}}{{Kewley}
  et~al.}{2006}]{kewley2006}
{Kewley} L.~J.,  {Geller} M.~J.,   {Barton} E.~J.,  2006, \mn@doi [\aj]
  {10.1086/500295}, \href {http://adsabs.harvard.edu/abs/2006AJ....131.2004K}
  {131, 2004}

\bibitem[\protect\citeauthoryear{{Lotz}, {Jonsson}, {Cox}  \& {Primack}}{{Lotz}
  et~al.}{2008}]{lotz2008}
{Lotz} J.~M.,  {Jonsson} P.,  {Cox} T.~J.,   {Primack} J.~R.,  2008, \mn@doi
  [\mnras] {10.1111/j.1365-2966.2008.14004.x}, \href
  {http://adsabs.harvard.edu/abs/2008MNRAS.391.1137L} {391, 1137}

\bibitem[\protect\citeauthoryear{{Maccagni}, {Morganti}, {Oosterloo},
  {Ger{\'e}b}  \& {Maddox}}{{Maccagni} et~al.}{2017}]{maccagni2017}
{Maccagni} F.~M.,  {Morganti} R.,  {Oosterloo} T.~A.,  {Ger{\'e}b} K.,
  {Maddox} N.,  2017, \mn@doi [\aap] {10.1051/0004-6361/201730563}, \href
  {http://adsabs.harvard.edu/abs/2017A%26A...604A..43M} {604, A43}

\bibitem[\protect\citeauthoryear{{Mihos} \& {Hernquist}}{{Mihos} \&
  {Hernquist}}{1996}]{mihos1996}
{Mihos} J.~C.,  {Hernquist} L.,  1996, \mn@doi [\apj] {10.1086/177353}, \href
  {http://adsabs.harvard.edu/abs/1996ApJ...464..641M} {464, 641}

\bibitem[\protect\citeauthoryear{{Mirabel} \& {Sanders}}{{Mirabel} \&
  {Sanders}}{1988}]{mirabel1988}
{Mirabel} I.~F.,  {Sanders} D.~B.,  1988, \mn@doi [\apj] {10.1086/166909},
  \href {https://ui.adsabs.harvard.edu/abs/1988ApJ...335..104M} {335, 104}

\bibitem[\protect\citeauthoryear{{Moreno}, {Torrey}, {Ellison}, {Patton},
  {Bluck}, {Bansal}  \& {Hernquist}}{{Moreno} et~al.}{2015}]{moreno2015}
{Moreno} J.,  {Torrey} P.,  {Ellison} S.~L.,  {Patton} D.~R.,  {Bluck}
  A.~F.~L.,  {Bansal} G.,   {Hernquist} L.,  2015, \mn@doi [\mnras]
  {10.1093/mnras/stv094}, \href
  {http://adsabs.harvard.edu/abs/2015MNRAS.448.1107M} {448, 1107}

\bibitem[\protect\citeauthoryear{{Moreno} et~al.,}{{Moreno}
  et~al.}{2019}]{moreno2019}
{Moreno} J.,  et~al., 2019, \mn@doi [\mnras] {10.1093/mnras/stz417}, \href
  {http://adsabs.harvard.edu/abs/2019MNRAS.485.1320M} {485, 1320}

\bibitem[\protect\citeauthoryear{{Morganti}, {Peck}, {Oosterloo}, {van
  Moorsel}, {Capetti}, {Fanti}, {Parma}  \& {de Ruiter}}{{Morganti}
  et~al.}{2009}]{morganti2009}
{Morganti} R.,  {Peck} A.~B.,  {Oosterloo} T.~A.,  {van Moorsel} G.,  {Capetti}
  A.,  {Fanti} R.,  {Parma} P.,   {de Ruiter} H.~R.,  2009, \mn@doi [\aap]
  {10.1051/0004-6361/200912605}, \href
  {http://adsabs.harvard.edu/abs/2009A%26A...505..559M} {505, 559}

\bibitem[\protect\citeauthoryear{{Moshir} et~al.,}{{Moshir}
  et~al.}{1990}]{moshir1990}
{Moshir} M.,  et~al., 1990, in Bulletin of the American Astronomical Society.
  p.~1325

\bibitem[\protect\citeauthoryear{{Mundell}, {Ferruit}  \& {Pedlar}}{{Mundell}
  et~al.}{2001}]{mundell2001}
{Mundell} C.~G.,  {Ferruit} P.,   {Pedlar} A.,  2001, \mn@doi [\apj]
  {10.1086/322508}, \href {http://adsabs.harvard.edu/abs/2001ApJ...560..168M}
  {560, 168}

\bibitem[\protect\citeauthoryear{{Pan} et~al.,}{{Pan} et~al.}{2018}]{pan2018}
{Pan} H.-A.,  et~al., 2018, \mn@doi [\apj] {10.3847/1538-4357/aaeb92}, \href
  {http://adsabs.harvard.edu/abs/2018ApJ...868..132P} {868, 132}

\bibitem[\protect\citeauthoryear{{Salim} et~al.,}{{Salim}
  et~al.}{2007}]{salim2007}
{Salim} S.,  et~al., 2007, \mn@doi [\apjs] {10.1086/519218}, \href
  {http://cdsads.u-strasbg.fr/abs/2007ApJS..173..267S} {173, 267}

\bibitem[\protect\citeauthoryear{{Sanders} \& {Mirabel}}{{Sanders} \&
  {Mirabel}}{1996}]{sanders1996}
{Sanders} D.~B.,  {Mirabel} I.~F.,  1996, \mn@doi [\araa]
  {10.1146/annurev.astro.34.1.749}, \href
  {http://adsabs.harvard.edu/abs/1996ARA%26A..34..749S} {34, 749}

\bibitem[\protect\citeauthoryear{{Satyapal}, {Ellison}, {McAlpine}, {Hickox},
  {Patton}  \& {Mendel}}{{Satyapal} et~al.}{2014}]{satyapal2014}
{Satyapal} S.,  {Ellison} S.~L.,  {McAlpine} W.,  {Hickox} R.~C.,  {Patton}
  D.~R.,   {Mendel} J.~T.,  2014, \mn@doi [\mnras] {10.1093/mnras/stu650},
  \href {http://adsabs.harvard.edu/abs/2014MNRAS.441.1297S} {441, 1297}

\bibitem[\protect\citeauthoryear{{Satyapal} et~al.,}{{Satyapal}
  et~al.}{2017}]{satyapal2017}
{Satyapal} S.,  et~al., 2017, \mn@doi [\apj] {10.3847/1538-4357/aa88ca}, \href
  {https://ui.adsabs.harvard.edu/abs/2017ApJ...848..126S} {848, 126}

\bibitem[\protect\citeauthoryear{{Schulz}, {Morganti}, {Nyland}, {Paragi},
  {Mahony}  \& {Oosterloo}}{{Schulz} et~al.}{2018}]{schulz2018}
{Schulz} R.,  {Morganti} R.,  {Nyland} K.,  {Paragi} Z.,  {Mahony} E.~K.,
  {Oosterloo} T.,  2018, \mn@doi [\aap] {10.1051/0004-6361/201833108}, \href
  {https://ui.adsabs.harvard.edu/abs/2018A&A...617A..38S} {617, A38}

\bibitem[\protect\citeauthoryear{{Scudder}, {Ellison}, {Torrey}, {Patton}  \&
  {Mendel}}{{Scudder} et~al.}{2012}]{scudder2012}
{Scudder} J.~M.,  {Ellison} S.~L.,  {Torrey} P.,  {Patton} D.~R.,   {Mendel}
  J.~T.,  2012, \mn@doi [\mnras] {10.1111/j.1365-2966.2012.21749.x}, \href
  {http://adsabs.harvard.edu/abs/2012MNRAS.426..549S} {426, 549}

\bibitem[\protect\citeauthoryear{{Srianand}, {Gupta}, {Momjian}  \&
  {Vivek}}{{Srianand} et~al.}{2015}]{srianand2015}
{Srianand} R.,  {Gupta} N.,  {Momjian} E.,   {Vivek} M.,  2015, \mn@doi
  [\mnras] {10.1093/mnras/stv1004}, \href
  {http://adsabs.harvard.edu/abs/2015MNRAS.451..917S} {451, 917}

\bibitem[\protect\citeauthoryear{{Tacconi}, {Genzel}, {Tecza}, {Gallimore},
  {Downes}  \& {Scoville}}{{Tacconi} et~al.}{1999}]{tacconi1999}
{Tacconi} L.~J.,  {Genzel} R.,  {Tecza} M.,  {Gallimore} J.~F.,  {Downes} D.,
  {Scoville} N.~Z.,  1999, \mn@doi [\apj] {10.1086/307839}, \href
  {http://adsabs.harvard.edu/abs/1999ApJ...524..732T} {524, 732}

\bibitem[\protect\citeauthoryear{{Takeuchi}, {Buat}, {Heinis}, {Giovannoli},
  {Yuan}, {Iglesias-P{\'a}ramo}, {Murata}  \& {Burgarella}}{{Takeuchi}
  et~al.}{2010}]{takeuchi2010}
{Takeuchi} T.~T.,  {Buat} V.,  {Heinis} S.,  {Giovannoli} E.,  {Yuan} F.-T.,
  {Iglesias-P{\'a}ramo} J.,  {Murata} K.~L.,   {Burgarella} D.,  2010, \mn@doi
  [\aap] {10.1051/0004-6361/200913476}, \href
  {http://adsabs.harvard.edu/abs/2010A%26A...514A...4T} {514, A4}

\bibitem[\protect\citeauthoryear{{Teng}, {Veilleux}  \& {Baker}}{{Teng}
  et~al.}{2013}]{teng2013}
{Teng} S.~H.,  {Veilleux} S.,   {Baker} A.~J.,  2013, \mn@doi [\apj]
  {10.1088/0004-637X/765/2/95}, \href
  {http://adsabs.harvard.edu/abs/2013ApJ...765...95T} {765, 95}

\bibitem[\protect\citeauthoryear{{Torrey}, {Cox}, {Kewley}  \&
  {Hernquist}}{{Torrey} et~al.}{2012}]{torrey2012}
{Torrey} P.,  {Cox} T.~J.,  {Kewley} L.,   {Hernquist} L.,  2012, \mn@doi
  [\apj] {10.1088/0004-637X/746/1/108}, \href
  {http://adsabs.harvard.edu/abs/2012ApJ...746..108T} {746, 108}

\bibitem[\protect\citeauthoryear{{Vermeulen} et~al.,}{{Vermeulen}
  et~al.}{2003}]{vermeulen2003}
{Vermeulen} R.~C.,  et~al., 2003, \mn@doi [\aap] {10.1051/0004-6361:20030468},
  \href {http://adsabs.harvard.edu/abs/2003A%26A...404..861V} {404, 861}

\bibitem[\protect\citeauthoryear{{Violino}, {Ellison}, {Sargent}, {Coppin},
  {Scudder}, {Mendel}  \& {Saintonge}}{{Violino} et~al.}{2018}]{violino2018}
{Violino} G.,  {Ellison} S.~L.,  {Sargent} M.,  {Coppin} K.~E.~K.,  {Scudder}
  J.~M.,  {Mendel} T.~J.,   {Saintonge} A.,  2018, \mn@doi [\mnras]
  {10.1093/mnras/sty345}, \href
  {http://adsabs.harvard.edu/abs/2018MNRAS.476.2591V} {476, 2591}

\bibitem[\protect\citeauthoryear{{Weinmann}, {van den Bosch}, {Yang}  \&
  {Mo}}{{Weinmann} et~al.}{2006}]{weinmann2006}
{Weinmann} S.~M.,  {van den Bosch} F.~C.,  {Yang} X.,   {Mo} H.~J.,  2006,
  \mn@doi [\mnras] {10.1111/j.1365-2966.2005.09865.x}, \href
  {http://adsabs.harvard.edu/abs/2006MNRAS.366....2W} {366, 2}

\bibitem[\protect\citeauthoryear{{Weston}, {McIntosh}, {Brodwin}, {Mann},
  {Cooper}, {McConnell}  \& {Nielsen}}{{Weston} et~al.}{2017}]{weston2017}
{Weston} M.~E.,  {McIntosh} D.~H.,  {Brodwin} M.,  {Mann} J.,  {Cooper} A.,
  {McConnell} A.,   {Nielsen} J.~L.,  2017, \mn@doi [\mnras]
  {10.1093/mnras/stw2620}, \href
  {http://adsabs.harvard.edu/abs/2017MNRAS.464.3882W} {464, 3882}

\bibitem[\protect\citeauthoryear{{White}, {Becker}, {Helfand}  \&
  {Gregg}}{{White} et~al.}{1997}]{white1997}
{White} R.~L.,  {Becker} R.~H.,  {Helfand} D.~J.,   {Gregg} M.~D.,  1997, \apj,
  \href {http://adsabs.harvard.edu/abs/1997ApJ...475..479W} {475, 479}

\bibitem[\protect\citeauthoryear{{York} et~al.,}{{York}
  et~al.}{2000}]{york2000}
{York} D.~G.,  et~al., 2000, \mn@doi [\aj] {10.1086/301513}, \href
  {http://adsabs.harvard.edu/abs/2000AJ....120.1579Y} {120, 1579}

\bibitem[\protect\citeauthoryear{{de Gasperin}, {Intema}  \& {Frail}}{{de
  Gasperin} et~al.}{2018}]{degasperin2018}
{de Gasperin} F.,  {Intema} H.~T.,   {Frail} D.~A.,  2018, \mn@doi [\mnras]
  {10.1093/mnras/stx3125}, \href
  {http://adsabs.harvard.edu/abs/2018MNRAS.474.5008D} {474, 5008}

\bibitem[\protect\citeauthoryear{{van Gorkom}, {Knapp}, {Ekers}, {Ekers},
  {Laing}  \& {Polk}}{{van Gorkom} et~al.}{1989}]{vangorkom1989}
{van Gorkom} J.~H.,  {Knapp} G.~R.,  {Ekers} R.~D.,  {Ekers} D.~D.,  {Laing}
  R.~A.,   {Polk} K.~S.,  1989, \mn@doi [\aj] {10.1086/115016}, \href
  {http://adsabs.harvard.edu/abs/1989AJ.....97..708V} {97, 708}

\makeatother
\end{thebibliography}
%
%
\begin{appendices}
\appendix
\setcounter{table}{0}
\renewcommand{\table}{A\arabic{table}}
\setcounter{figure}{0}
\renewcommand{\figure}{A\arabic{figure}}
\section{Properties of the full sample of mergers}
\label{appendix2}
%
\begin{table*} 
\caption{Properties of the full sample of mergers.}
\centering
\begin{small}
\begin{tabular}{p{3.2cm}p{0.6cm}p{0.5cm}p{0.6cm}p{0.6cm}p{0.6cm}p{0.6cm}p{0.6cm}p{0.6cm}p{0.6cm}p{1.0cm}p{1.0cm}p{1.0cm}p{0.6cm}}
\hline
Coordinates & $z$ & $\rho$ & log        & $\alpha_{0.15}^{1.4}$ & log         & $M_*$ & log         & SFR         & Stage & \nhi\      & FWHM   & \vshift\ & Ref. \\
(J2000)     &     &        & $P_{1.4}$  &                       & $M_*$       & ratio & \lir\       &             &       &            &        &          &      \\
            &     & (kpc)  & (W         &                       & (M$_\odot$) &       & (L$_\odot$) & (M$_\odot$  &       & ($10^{20}$ & (\kms) & (\kms)   &      \\
            &     &        & Hz$^{-1}$) &                       &             &       &             & ~yr$^{-1}$) &       & \cms)      &        &          &      \\
(1)         & (2) & (3)    & (4)        & (5)                   & (6)         & (7)   & (8)         & (9)         & (10)  & (11)       & (12)   & (13)     & (14) \\
\hline
00:54:03.99 $+$73:05:05.40 & 0.01570 & ~~5  & 22.70 & $-$0.53 &   --- & --- & 11.48 &   ---  & pr & 37     & 183 & 84     & 1  \\
01:20:02.72 $+$14:21:42.94 & 0.03120 & ~~33 & 22.96 & $-$0.29 &  9.87 & 2.7 & 11.73 &  5.799 & pr & 60     & 191 & 5      & 2  \\
08:38:24.01 $+$25:45:16.28 & 0.01818 & ~~0  & 22.83 & $-$0.24 & 10.36 & --- & 11.58 &  0.300 & po & 67     & 108 & 64     & 3  \\
09:04:34.99 $+$14:35:38.18 & 0.04952 & ~~7  & 23.45 & $-$0.74 & 10.78 & 3.5 & 11.84 &  6.605 & on & $\le$7 & --- & ---    & 2  \\ 
09:15:55.51 $+$44:19:58.00 & 0.03979 & ~~9  & 23.06 & $-$0.22 & 10.81 & 1.6 & 11.65 &  1.628 & on & 47     & 258 & 34     & 2  \\
09:35:51.59 $+$61:21:11.33 & 0.03939 & ~~0  & 23.70 & $-$0.28 & 10.96 & --- & 12.01 &  0.298 & po & 54     & 536 & $-$78  & 4  \\
09:42:21.98 $+$06:23:35.23 & 0.12368 & ~~5  & 24.62 & ---     & 10.44 & --- &   --- &    --- & on & 91     & 75  & 0      & 5  \\
09:45:42.05 $-$14:19:34.98 & 0.00771 & ~~28 & 22.48 & $-$0.63 &   --- & --- & 10.47 &    --- & pr & 57     & 95  & $-$18  & 6  \\
10:36:31.96 $+$02:21:45.89 & 0.04990 & ~~2  & 24.00 & $-$0.64 & 10.40 & --- & 12.08 &  2.688 & po & 92     & 289 & $-$6   & 1  \\
10:53:27.25 $+$20:58:35.92 & 0.05264 & ~~6  & 23.72 & $-$0.12 & 10.75 & 1.6 &   --- &  0.112 & pr & 5      & 156 & $-$58  & 4  \\
11:00:17.98 $+$10:02:56.84 & 0.03624 & ~~13 & 23.59 & $-$0.40 & 10.67 & 0.7 & 11.43 &  0.568 & pr & 121    & 61  & $-$49  & 1  \\
11:03:53.95 $+$40:50:59.91 & 0.03479 & ~~12 & 22.94 & $-$0.38 & 10.26 & 23  & 11.61 &  0.786 & pr & 148    & 270 & 107    & 2  \\ 
11:08:26.51 $-$10:15:21.70 & 0.02730 & ~~6  & 23.72 & $-$0.58 &   --- & --- &   --- &    --- & pr & $\le$6 & --- & ---    & 1  \\	  
11:28:33.41 $+$58:33:46.20 & 0.01046 & ~~4  & 22.62 & $-$0.59 & 10.36 & 1.6 & 11.79 &  0.004 & on & 94     & 88  & 165    & 3  \\ 
12:14:18.25 $+$29:31:46.70 & 0.06326 & ~~8  & 23.93 & $-$0.50 & 11.02 & 21  & 11.41 &  0.243 & on & $\le$2 & --- & ---    & 1  \\    
12:56:14.22 $+$56:52:25.27 & 0.04217 & ~~0  & 23.99 & $-$0.38 & 10.78 & --- & 12.56 &    --- & po & 29     & 179 & 0      & 7  \\ 
13:01:25.26 $+$29:18:49.53 & 0.02340 & ~~10 & 22.68 & $-$0.22 & 10.59 & 0.3 & 11.31 &  0.952 & on & 15     & 148 & 54     & 8  \\ 
13:01:50.29 $+$04:20:00.54 & 0.03736 & ~~0  & 22.96 & $-$0.57 & 10.84 & --- & 11.76 &  0.629 & po & 49     & 299 & $-$3   & 2  \\ 
13:15:35.10 $+$62:07:28.43 & 0.03083 & ~~21 & 23.02 & $-$0.35 & 10.31 & 0.8 & 11.75 &  5.374 & on & 55     & 291 & 93     & 1  \\
13:20:35.40 $+$34:08:21.75 & 0.02306 & ~~0  & 23.12 & $-$0.30 & 10.43 & --- & 11.67 &  0.924 & po & 113    & 197 & 73     & 1  \\
13:34:55.94 $+$13:44:31.74 & 0.02312 & ~~12 & 22.59 & $-$0.07 & 11.12 & 14  &   --- &  0.053 & pr & 29     & 130 & $-$101 & 8  \\
13:38:17.27 $+$48:16:32.20 & 0.02758 & ~~6  & 22.78 & $-$0.68 & 10.61 & 1.0 & 11.54 &  2.015 & on & 13     & 140 & 176    & 8  \\ 
13:44:42.16 $+$55:53:13.53 & 0.03734 & ~~0  & 23.59 & $-$0.48 & 10.82 & --- & 12.19 &  4.778 & po & 86     & 570 & 85     & 4  \\ 
13:47:33.36 $+$12:17:24.27 & 0.12047 & ~~0  & 26.26 & $-$0.18 & 11.59 & --- & 12.41 &  0.188 & po & 3      & 135 & 45     & 9  \\ 
13:56:02.63 $+$18:22:17.68 & 0.05036 & ~~4  & 24.40 & $-$0.77 & 11.01 & --- & 11.82 &  9.223 & on & $\le$1 & --- & ---    & 1  \\   
13:56:46.12 $+$10:26:09.09 & 0.12313 & ~~2  & 24.50 & ---     & 11.35 & --- & 12.29 &  8.464 & po & 47     & 108 & 148    & 1  \\ 
14:41:04.37 $+$53:20:08.74 & 0.10502 & ~~54 & 24.03 & $-$0.74 & 10.87 & 2.3 & 12.22 &  7.004 & pr & $\le$5 & --- & ---    & 8  \\    
14:57:00.68 $+$24:37:03.55 & 0.03367 & ~~26 & 23.24 & $-$0.69 & 10.75 & 1.7 & 11.68 &  6.256 & pr & 22     & 44  & 118    & 2  \\ 
15:08:05.59 $+$34:23:22.87 & 0.04557 & ~~0  & 23.77 & $-$0.25 & 10.62 & --- & 11.58 &  0.040 & po & 125    & 102 & $-$178 & 10 \\
15:18:06.13 $+$42:44:45.01 & 0.04036 & ~~6  & 23.30 & $-$0.57 & 10.62 & 1.7 & 11.92 & 14.726 & on & 47     & 143 & $-$2   & 2  \\ 
15:31:43.46 $+$24:04:19.17 & 0.09646 & ~~5  & 23.75 & ---     & 11.40 & 79  & 11.51 &  1.039 & on & 92     & 33  & 235    & 11 \\ 
15:34:57.20 $+$23:30:13.24 & 0.01840 & ~~0  & 23.35 & $-$0.15 & 10.77 & --- & 12.19 &  0.639 & po & 132    & 238 & $-$32  & 12 \\
16:04:26.51 $+$17:44:31.17 & 0.04089 & ~~24 & 23.43 & $-$0.18 & 11.08 & 1.0 & 11.17 &  0.045 & pr & $\le$3 & --- & ---    & 4  \\ 
16:05:43.25 $+$37:10:45.03 & 0.06649 & ~~0  & 23.37 & $-$0.56 & 10.47 & --- & 11.75 &  4.467 & po & 67     & 81  & $-$32  & 2  \\
16:38:03.65 $+$26:43:30.62 & 0.06521 & ~~6  & 23.54 & ---     & 10.72 & 1.4 &   --- &  0.008 & pr & 25     & 111 & 96     & 8  \\
16:52:58.89 $+$02:24:03.11 & 0.02448 & ~~0  & 23.72 & $-$0.72 & 11.08 & --- & 11.86 &    --- & po & 128    & 348 & $-$43  & 13 \\
20:54:49.61 $+$00:41:53.07 & 0.20276 & ~~11 & 25.64 & ---     & 10.78 & 3.8 &   --- &  0.962 & pr & 55     & 60  & 92     & 1  \\
20:57:23.90 $+$17:07:39.00 & 0.03593 & ~~6  & 22.74 & $-$0.67 & 10.31 & 3.3 & 11.95 &    --- & on & 89     & 100 & 144    & 2  \\
\hline
\end{tabular}
\label{tab:fullsample}
\begin{flushleft} {\it Notes.}
Column 1: source coordinates.
Column 2: systemic redshift of galaxy hosting the radio source, from optical emission lines. 
Column 3: projected separation between two merging galaxies. Zero separation indicates only single nuclei is discernible in SDSS images.
Column 4: radio power based on FIRST 1.4\,GHz fluxes.
Column 5: spectral index from TGSS 147~MHz and NVSS 1.4\,GHz fluxes.
Column 6: stellar mass from SDSS photometry using {\it kcorrect}.  
Column 7: stellar mass ratio for galaxy pairs (where the galaxy with the stronger radio emission is in the numerator).
Column 8: total infrared luminosity ($8-1000~\micron$) from IRAS or AKARI. 
Column 9: SFR from MPA-JHU SDSS DR7 spectroscopic catalog. 
Column 10: merger stage $-$ `pr': pre-merger, `on': ongoing, `po': post-merger.
Column 11: \nhi\ (3$\sigma$ upper limit per 100\,\kms) of detections (non-detections), for \ts\ = 100\,K and \fc\ = 1.
Column 12: full width at half the optical depth of the absorption. 
Column 13: velocity shift between systemic redshift and peak optical depth of \hi\ absorption. 
Column 14: references for \hi\ \21\ absorption $-$ 
1: D18; 2: this work; 3: \citet{dickey1986}; 4: \citet{gereb2015}; 5: \citet{srianand2015}; 6: \citet{gallimore1999}; 7: \citet{carilli1998b}; 
8: \citet{maccagni2017}; 9: \citet{gupta2006}; 10: \citet{chandola2011}; 11: \citet{chandola2012}; 12: \citet{mundell2001}; 13: \citet{baan2007} \\ 
\end{flushleft}
\end{small}
\end{table*}
%
\section{Details of the new sample of mergers}
\label{appendix1}
\begin{figure*}
\subfloat[J0120$+$1421 (pre-merger)]{\includegraphics[width=0.32\textwidth, bb=110 50 730 660, clip=true]{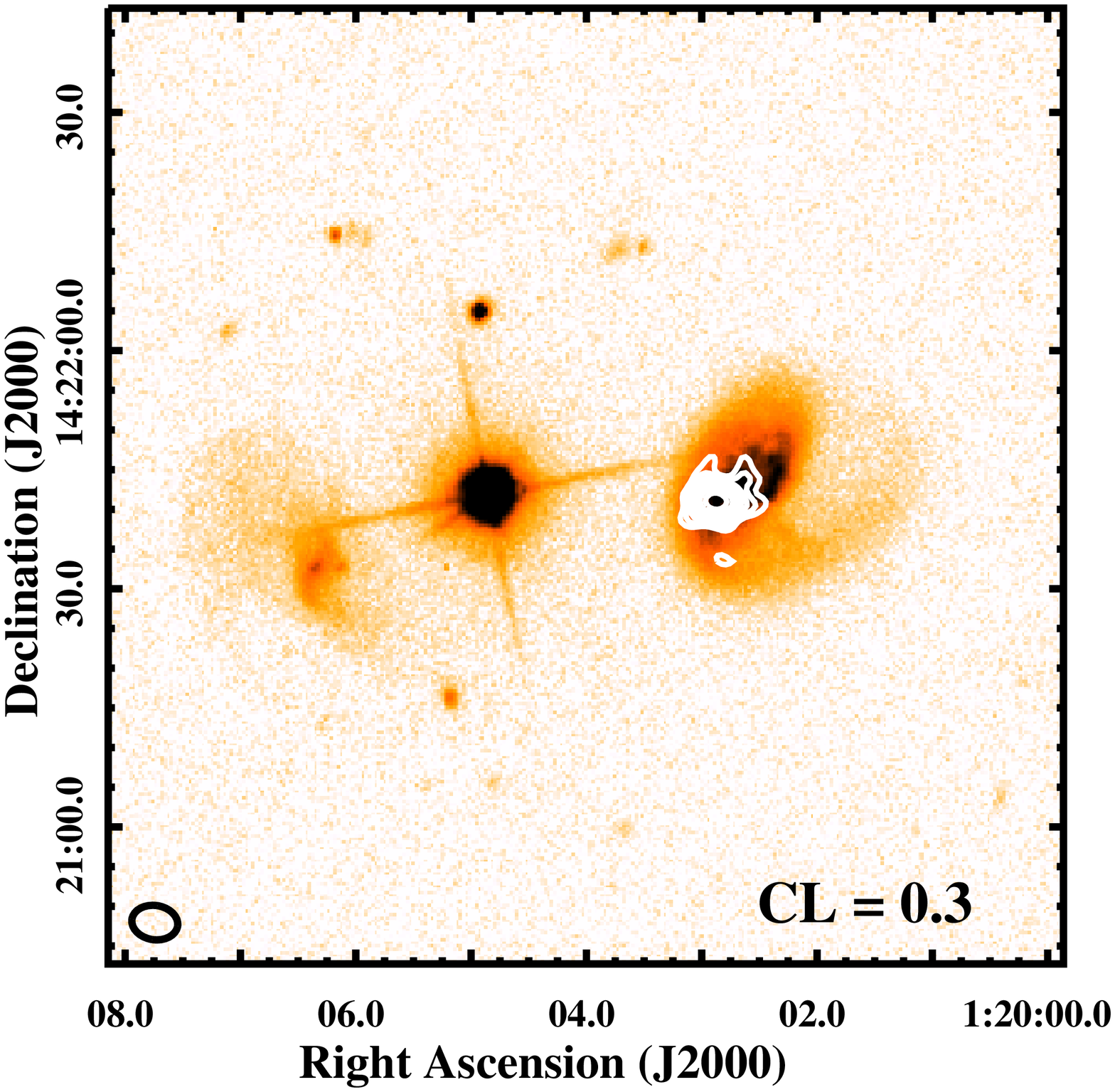} } 
\subfloat[J0904$+$1435 (ongoing)]{\includegraphics[width=0.31\textwidth, bb=120 40 740 670, clip=true]{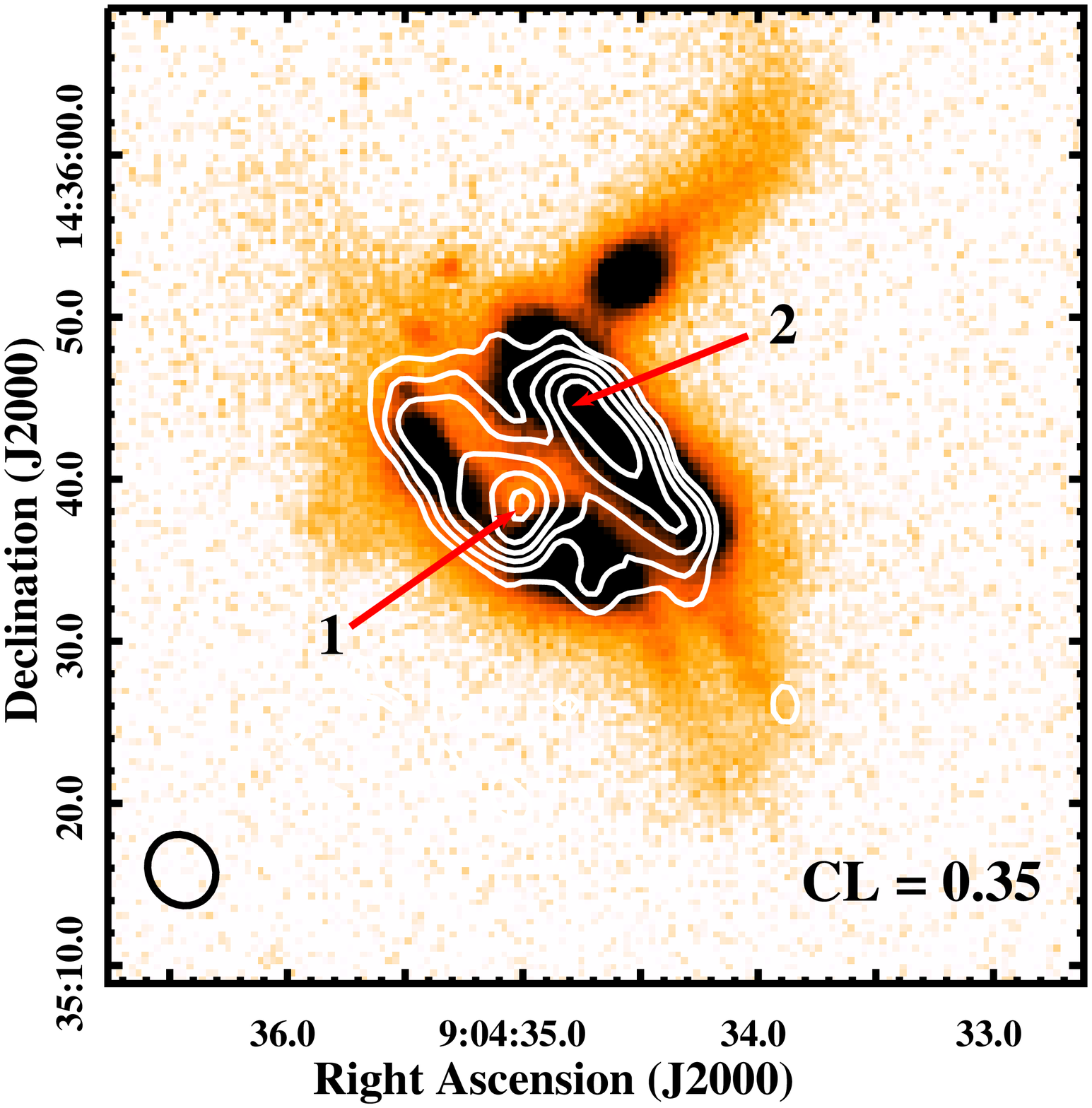} } 
\subfloat[J0915$+$4419 (ongoing)]{\includegraphics[width=0.31\textwidth, bb=100 30 745 680, clip=true]{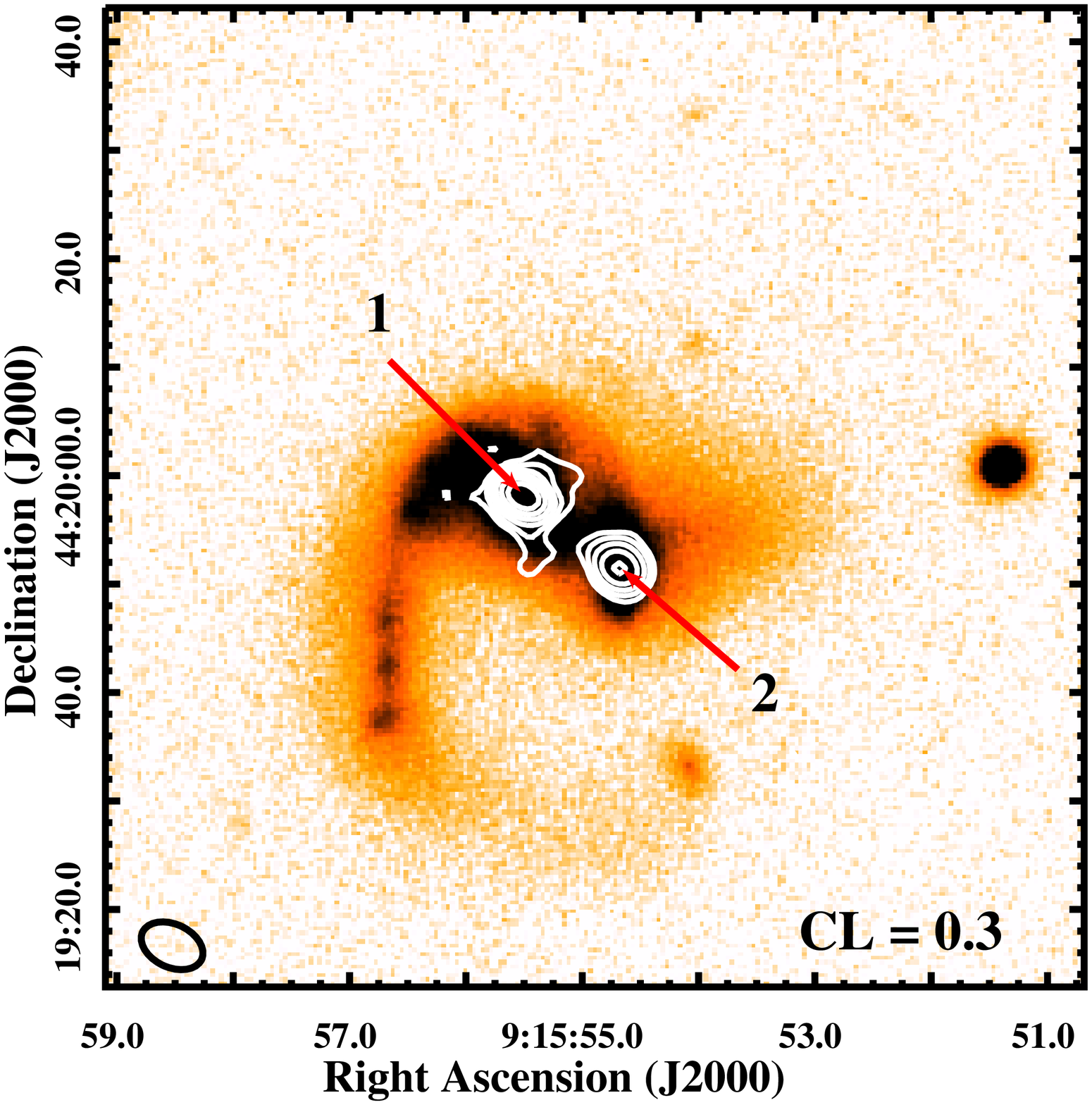} } \hspace{0.01cm}                     
\subfloat[J1103$+$4050 (pre-merger)]{\includegraphics[width=0.32\textwidth, bb=120 40 740 680, clip=true]{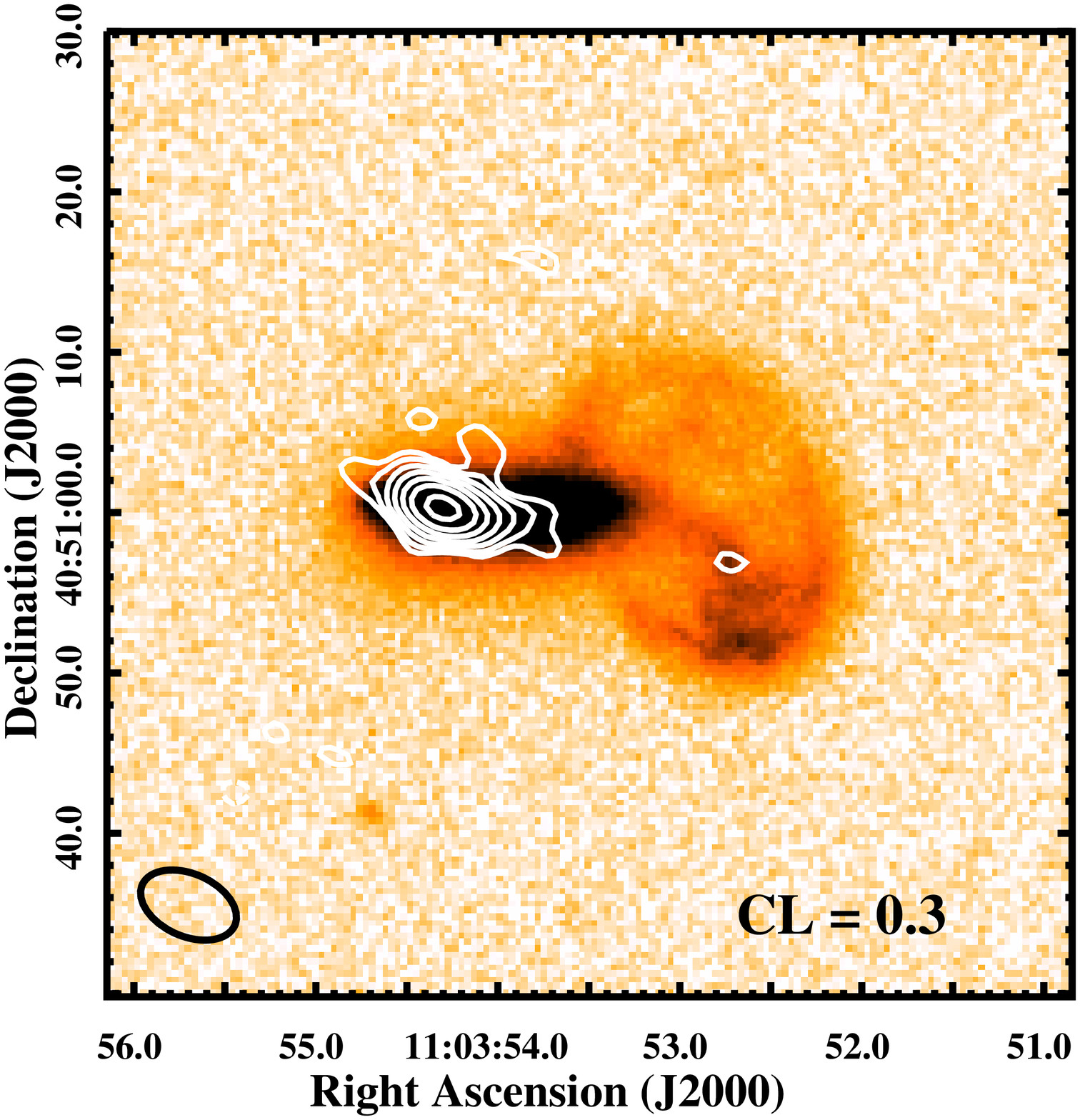} } 
\subfloat[J1301$+$0420 (post-merger)]{\includegraphics[width=0.32\textwidth, bb=75 25 720 680, clip=true]{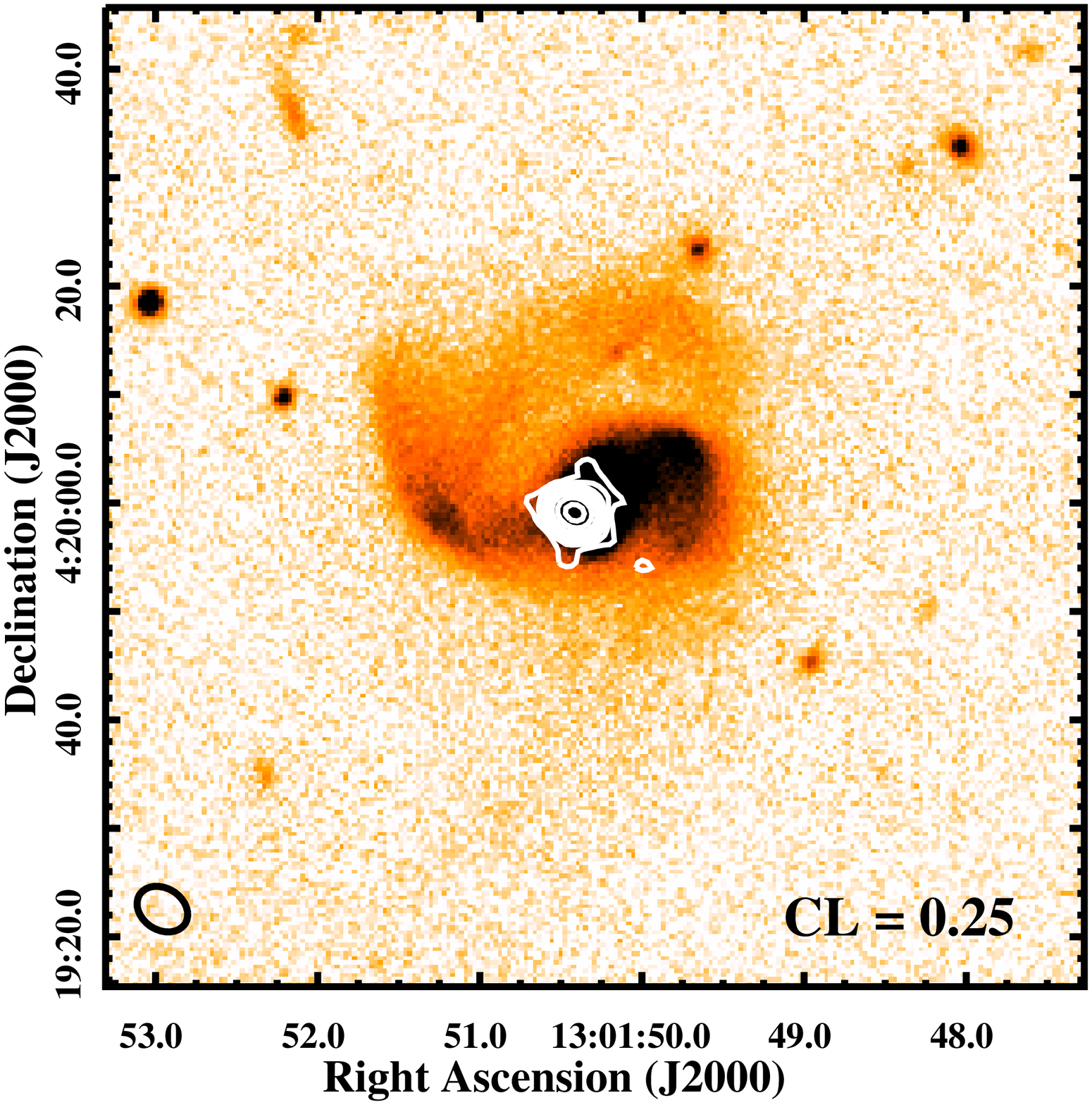} } 
\subfloat[J1457$+$2437 (pre-merger)]{\includegraphics[width=0.32\textwidth, bb=85 55 695 665, clip=true]{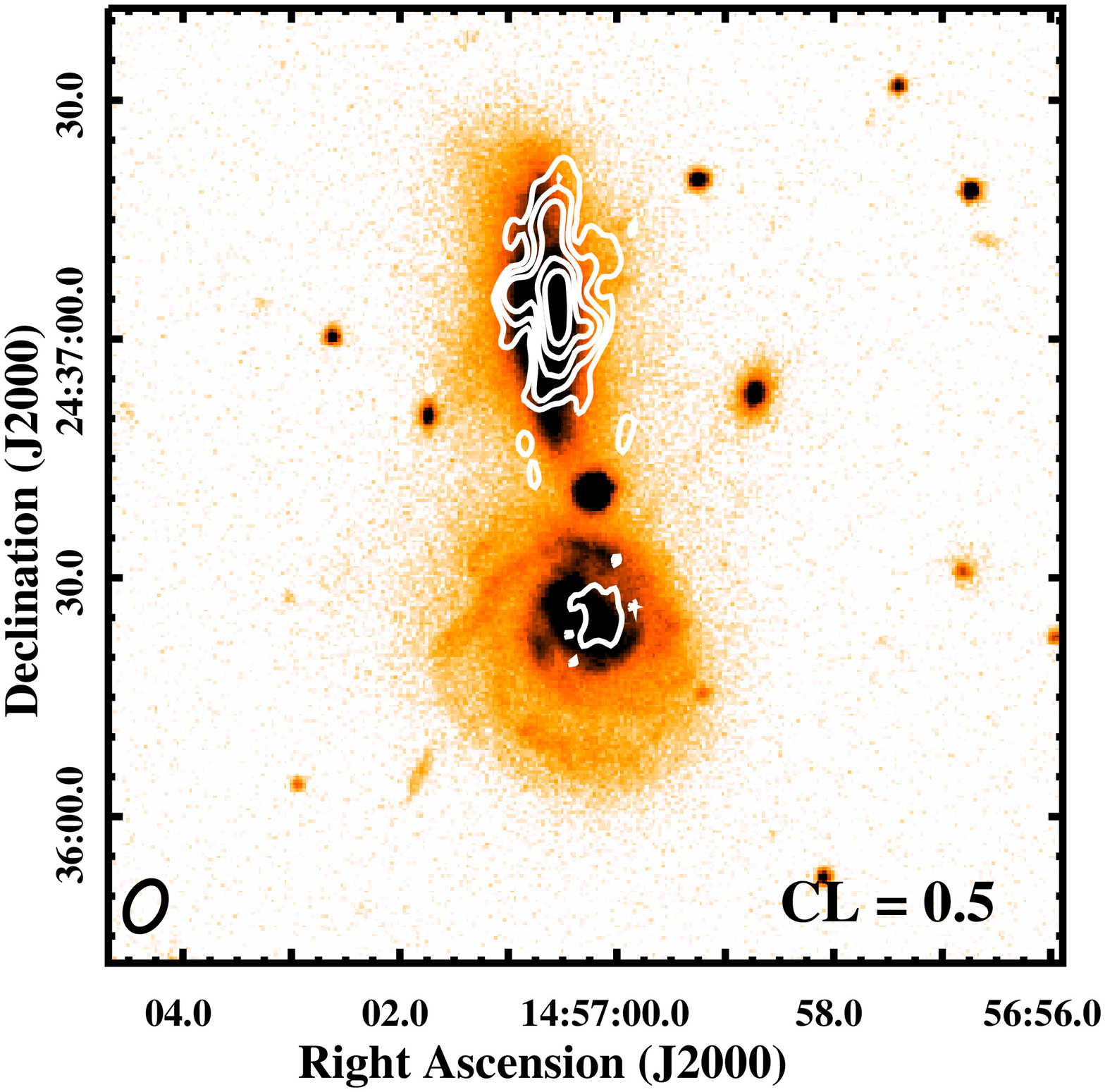} } \hspace{0.01cm}
\subfloat[J1518$+$4244 (ongoing)]{\includegraphics[width=0.32\textwidth, bb=60 40 720 690, clip=true]{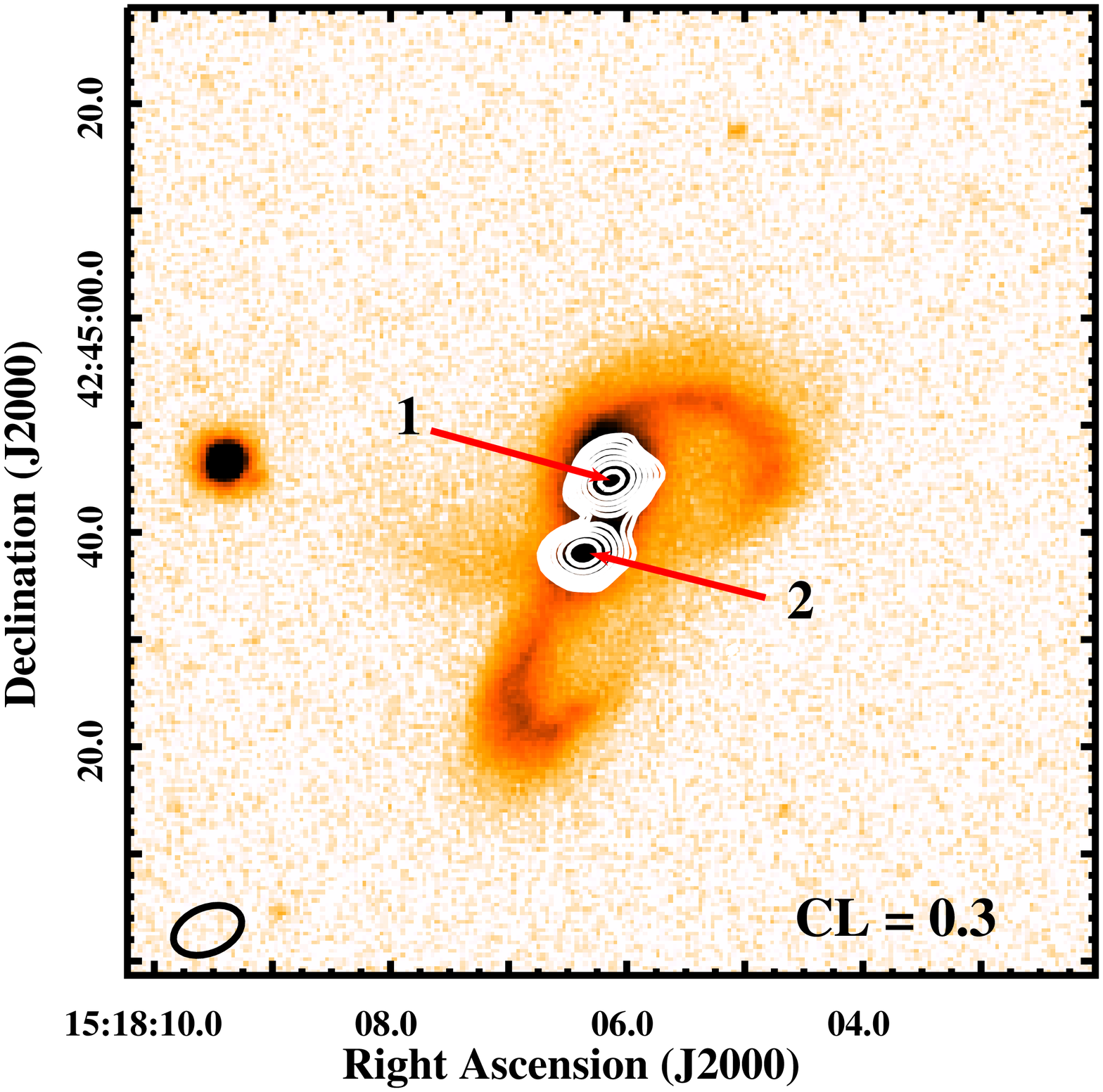} } 
\subfloat[J1605$+$3710 (post-merger)]{\includegraphics[width=0.32\textwidth, bb=70 50 695 680, clip=true]{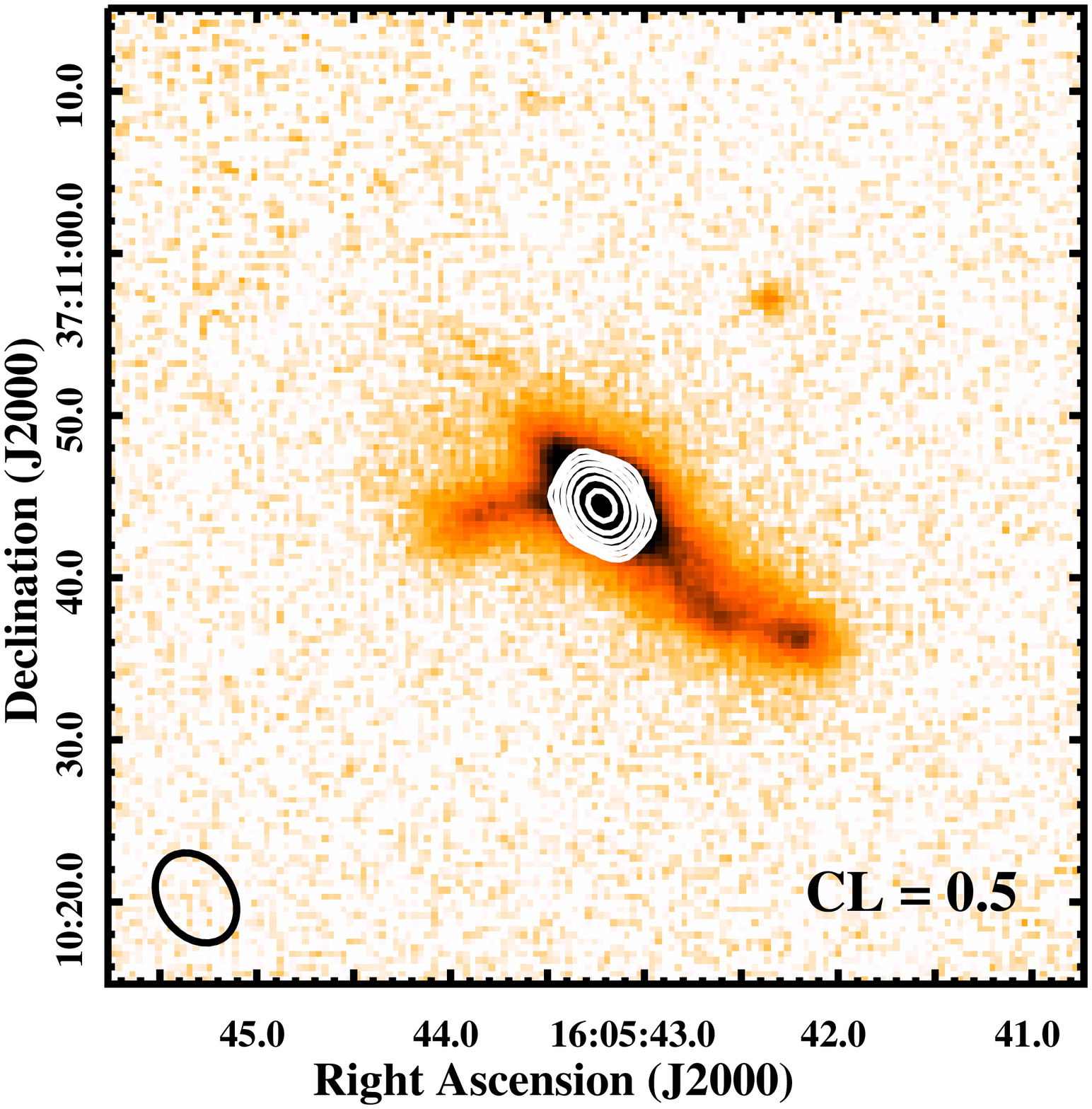} }
\subfloat[J2057$+$1707 (ongoing)]{\includegraphics[width=0.32\textwidth, bb=65 40 710 690, clip=true]{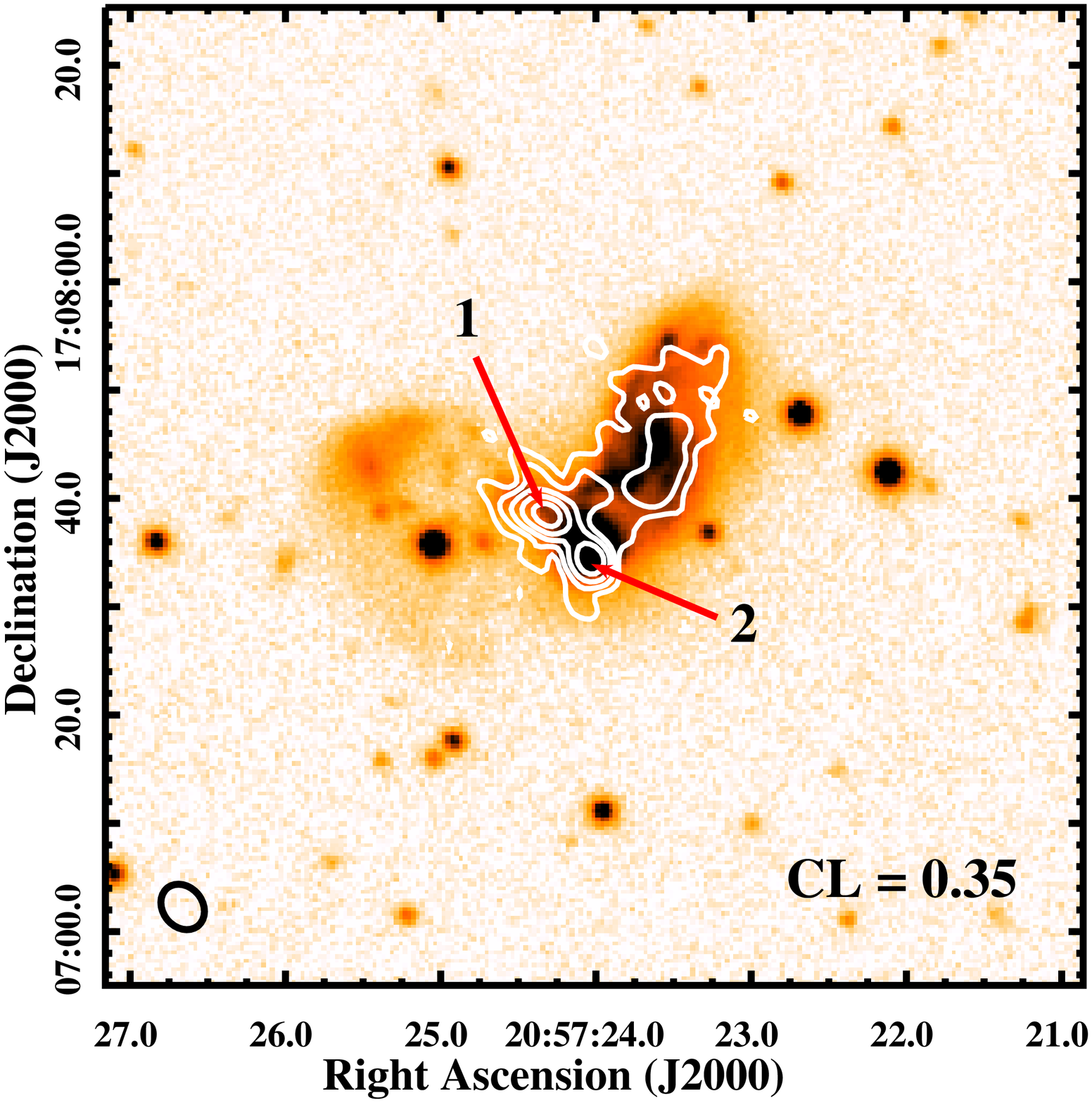} } 
\caption{SDSS $r$-band images of the new merger sample, overlaid with GMRT 1.4 GHz continuum contours. Restoring beam of the continuum map is 
shown in the bottom left corner. Contour levels are plotted as CL $\times$ ($-$1,1,2,4,8,...)\,\mjb, where CL is given in the bottom right corner. 
Contours start at 5 times the rms noise in the continuum map. Solid (dashed) lines correspond to positive (negative) values. Multiple continuum 
peaks are marked as listed in Table~\ref{tab:results}. The merger stage is indicated below each image.}
\label{fig:overlay}
\end{figure*}
\begin{figure*}
\subfloat{\includegraphics[width=0.7\textwidth, angle=90]{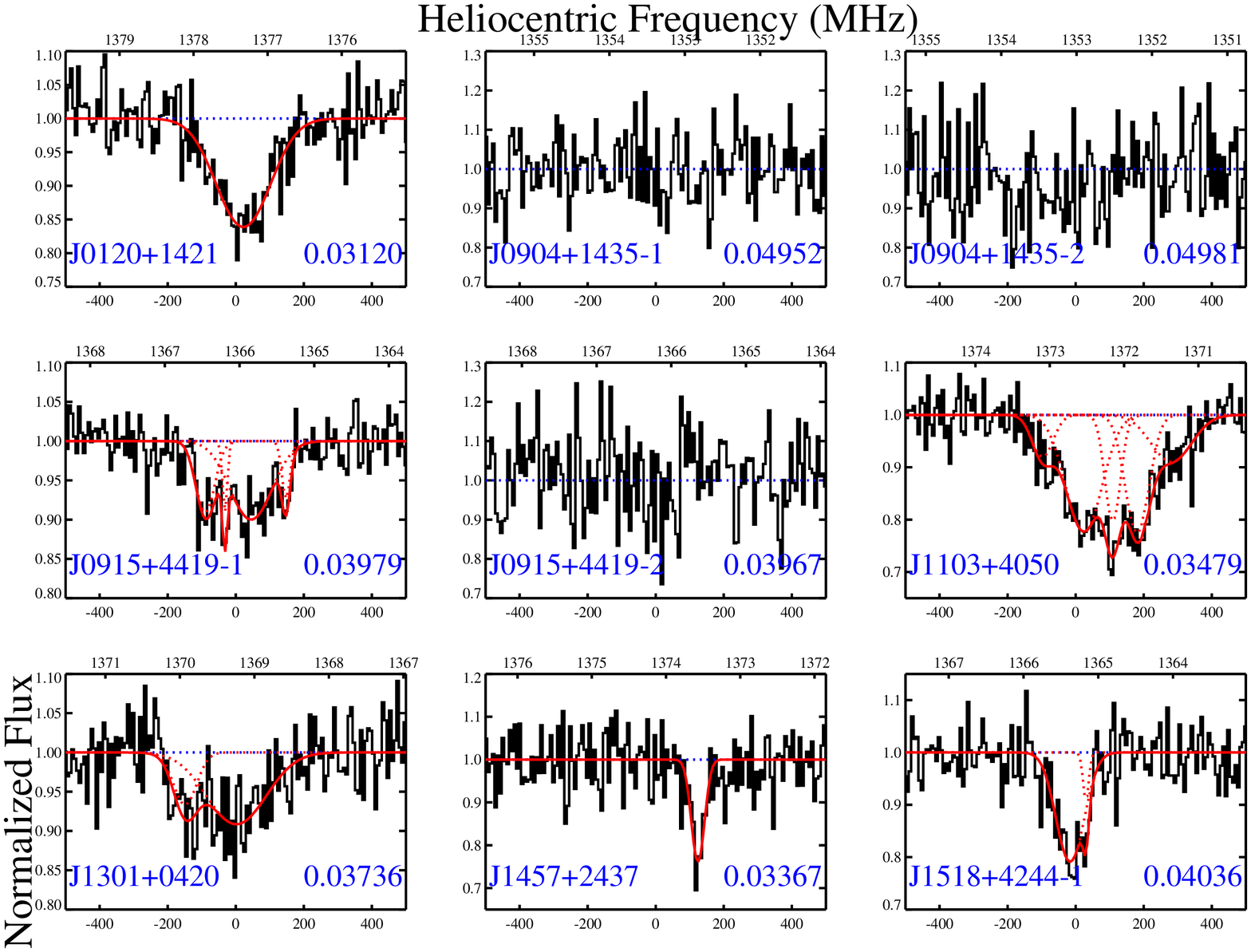} } \vskip-0.01cm
\subfloat{\includegraphics[width=0.45\textwidth, angle=90, bb=210 20 575 775, clip=true]{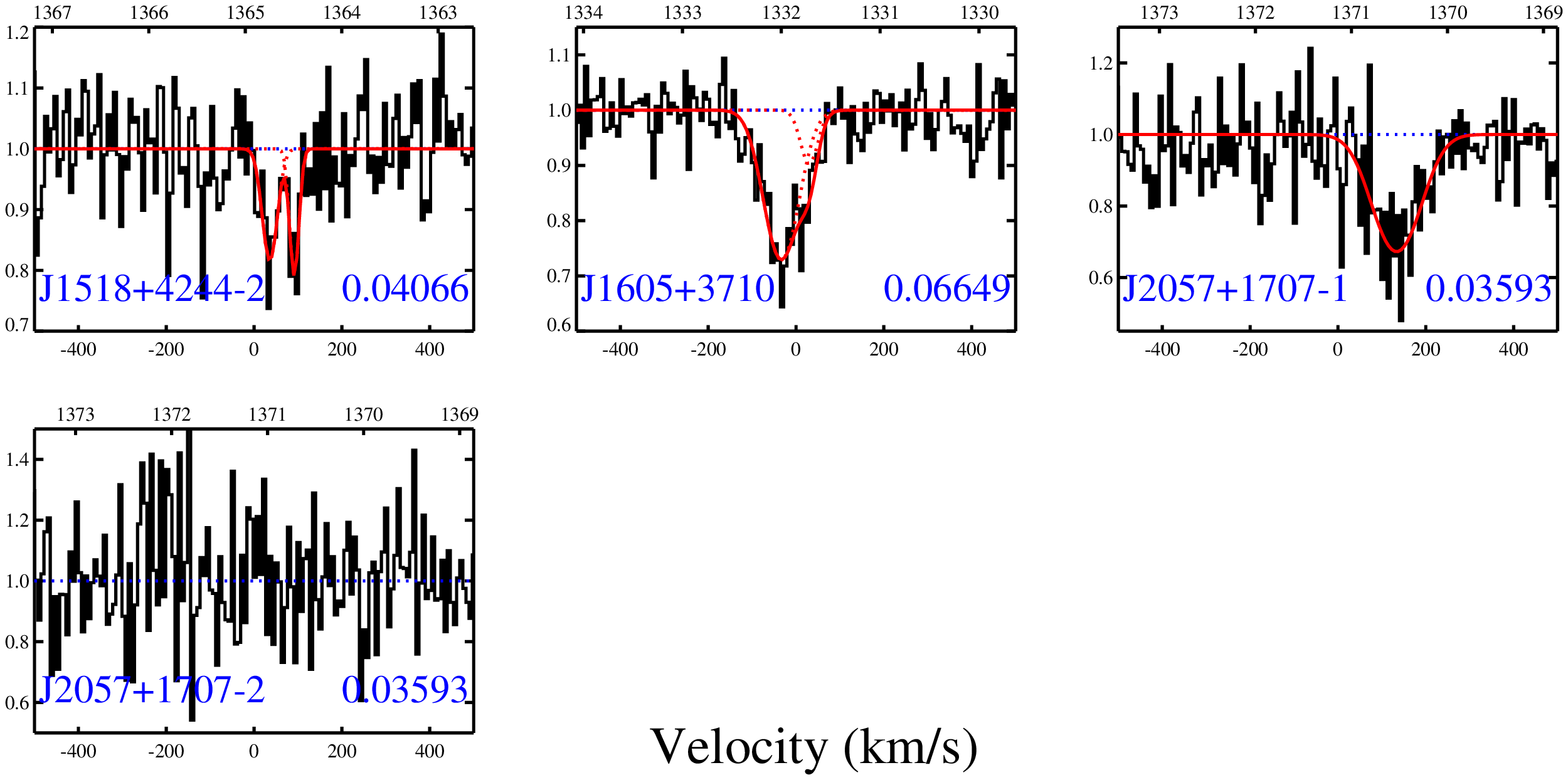} } 
\caption{Normalised \hi\ absorption spectra towards the radio sources in our new merger sample. The velocity scale is with respect 
to the systemic redshift as given in the bottom right of each plot. Gaussian fits are overplotted for detections (individual component 
in dotted line and total fit in solid line).}
\label{fig:21cmspectra}
\end{figure*}
\begin{table*}
\caption{\hi\ absorption properties of the new sample of mergers.}
\centering
\begin{tabular}{cccccccccc}
\hline
Source & $z$ & Peak     & Total    & Spectral & $\tau_{\rm p}$ & \taudv\ & \nhi\             & $v_{\rm 90}$ & \vshift\ \\
       &     & 1.4\,GHz & 1.4\,GHz & rms      &                &         & (\ts$/100$ K)     &              &          \\
       &     & Flux     & Flux     &          &                &         & ($1/$\fc)         &              &          \\
       &     & Density  & Density  &          &                &         &                   &              &          \\
       &     & (\mjb)   & (mJy)    & (\mjb)   &                & (\kms)  & ($10^{20}$\,\cms) & (\kms)       & (\kms)   \\
(1)    & (2) & (3)      & (4)      & (5)      & (6)            & (7)     & (8)               & (9)          & (10)     \\
\hline
J0120$+$1421   & 0.03120 & 33 & 41 & 1.1 & 0.24 $\pm$ 0.03 & 33 $\pm$ 2 & 60 $\pm$ 3  & 241 $\pm$ 4  & 5     $\pm$ 10 \\
J0904$+$1435-1 & 0.04952 & 13 & 49 & 1.0 & $\le$0.08       & $\le$4     & $\le$7      & ---          & ---            \\
J0904$+$1435-2 & 0.04981 & 10 & 45 & 1.0 & $\le$0.10       & $\le$8     & $\le$14     & ---          & ---            \\
J0915$+$4419-1 & 0.03979 & 18 & 22 & 0.5 & 0.16 $\pm$ 0.03 & 26 $\pm$ 1 & 47 $\pm$ 3  & 265 $\pm$ 13 & 34    $\pm$ 11 \\
J0915$+$4419-2 & 0.03967 &  5 &  9 & 0.5 & $\le$0.09       & $\le$8     & $\le$14     & ---          & ---            \\
J1103$+$4050   & 0.03479 & 26 & 31 & 1.0 & 0.37 $\pm$ 0.04 & 81 $\pm$ 2 & 148 $\pm$ 4 & 370 $\pm$ 22 & 107   $\pm$ 14 \\
J1301$+$0420   & 0.03736 & 22 & 28 & 0.8 & 0.17 $\pm$ 0.04 & 27 $\pm$ 2 & 49 $\pm$ 8  & 335 $\pm$ 14 & $-$3  $\pm$ 11 \\
J1457$+$2437   & 0.03367 & 16 & 66 & 1.0 & 0.36 $\pm$ 0.06 & 12 $\pm$ 2 & 22 $\pm$ 3  &  57 $\pm$ 1  & 118   $\pm$ 10 \\
J1518$+$4244-1 & 0.04036 & 26 & 33 & 1.2 & 0.28 $\pm$ 0.05 & 26 $\pm$ 2 & 47 $\pm$ 3  & 143 $\pm$ 4  & $-$2  $\pm$ 10 \\
J1518$+$4244-2 & 0.04066 & 15 & 20 & 1.1 & 0.31 $\pm$ 0.07 & 14 $\pm$ 2 & 26 $\pm$ 4  &  86 $\pm$ 2  & 33    $\pm$ 10 \\
J1605$+$3710   & 0.06649 & 22 & 22 & 1.1 & 0.44 $\pm$ 0.05 & 37 $\pm$ 2 & 67 $\pm$ 4  & 154 $\pm$ 6  & $-$32 $\pm$ 10 \\
J2057$+$1707-1 & 0.03593 &  9 & 19 & 1.0 & 0.74 $\pm$ 0.11 & 49 $\pm$ 4 & 89 $\pm$ 8  & 157 $\pm$ 2  & 144   $\pm$ 10 \\
J2057$+$1707-2 & 0.03593 &  5 & 10 & 0.9 & $\le$0.19       & $\le$13    & $\le$23     & ---          & ---            \\
\hline
\end{tabular}
\label{tab:results}
\begin{flushleft} 
{\it Notes.}
Column 1: galaxy merger name. In case of multiple continuum peaks, we number them, starting with 1 for the strongest.
Column 2: systemic redshift of galaxy hosting the radio source.
Column 3: peak 1.4\,GHz flux density.
Column 4: total 1.4\,GHz flux density. 
Column 5: spectral rms (at resolution of $\sim7$\,\kms). 
Column 6: peak optical depth (1$\sigma$ upper limit) of detections (non-detections).
Column 7: integrated optical depth (3$\sigma$ upper limit for 100\,\kms\ linewidth) of detections (non-detections).
Column 8: \nhi\ (3$\sigma$ upper limit) of detections (non-detections).
Column 9: velocity width which contains 90\% of the total optical depth for detections. 
Column 10: velocity shift between systemic redshift and velocity of peak optical depth for detections.
Positive (negative) sign indicates redshifted (blueshifted) absorption.
\end{flushleft}
\end{table*}
\begin{table*} 
\caption{Gaussian fit parameters to \hi\ absorption lines detected in the new sample of mergers.}
\centering
\begin{tabular}{cccc}
\hline
Source & $z$ & FWHM   & \taup\ \\
       &     & (\kms) &        \\
(1)    & (2) & (3)    & (4)    \\
\hline
J0120$+$1421   & 0.03128 $\pm$ 0.00001 & 181 $\pm$  3 & 0.18 $\pm$ 0.01 \\
J0915$+$4419-1 & 0.03948 $\pm$ 0.00001 &  65 $\pm$  5 & 0.10 $\pm$ 0.02 \\
               & 0.03968 $\pm$ 0.00001 &  17 $\pm$  2 & 0.09 $\pm$ 0.03 \\
               & 0.03995 $\pm$ 0.00001 & 143 $\pm$ 11 & 0.11 $\pm$ 0.02 \\
               & 0.04030 $\pm$ 0.00001 &  28 $\pm$  4 & 0.08 $\pm$ 0.02 \\
J1301$+$0420   & 0.03685 $\pm$ 0.00001 &  79 $\pm$  8 & 0.07 $\pm$ 0.02 \\
               & 0.03737 $\pm$ 0.00002 & 203 $\pm$ 12 & 0.10 $\pm$ 0.01 \\
J1457$+$2437   & 0.03410 $\pm$ 0.00001 &  44 $\pm$  1 & 0.27 $\pm$ 0.01 \\
J1518$+$4244-1 & 0.04021 $\pm$ 0.00001 &  93 $\pm$  2 & 0.23 $\pm$ 0.01 \\
               & 0.04037 $\pm$ 0.00001 &  19 $\pm$  3 & 0.10 $\pm$ 0.03 \\
J1518$+$4244-2 & 0.04049 $\pm$ 0.00001 &  39 $\pm$  1 & 0.20 $\pm$ 0.01 \\
               & 0.04106 $\pm$ 0.00001 &  24 $\pm$  1 & 0.23 $\pm$ 0.02 \\
J1605$+$3710   & 0.04062 $\pm$ 0.00001 &  90 $\pm$  3 & 0.32 $\pm$ 0.02 \\
               & 0.06660 $\pm$ 0.00001 &  46 $\pm$  5 & 0.10 $\pm$ 0.03 \\
J1103$+$4050   & 0.06615 $\pm$ 0.00001 &  72 $\pm$  7 & 0.08 $\pm$ 0.02 \\
               & 0.03487 $\pm$ 0.00001 & 121 $\pm$  9 & 0.25 $\pm$ 0.04 \\
               & 0.03517 $\pm$ 0.00001 &  52 $\pm$  4 & 0.23 $\pm$ 0.05 \\
               & 0.03541 $\pm$ 0.00001 &  79 $\pm$  6 & 0.26 $\pm$ 0.05 \\
               & 0.03576 $\pm$ 0.00004 & 137 $\pm$ 17 & 0.09 $\pm$ 0.03 \\
J2057$+$1707-1 & 0.03525 $\pm$ 0.00001 & 122 $\pm$  1 & 0.40 $\pm$ 0.01 \\
\hline
\end{tabular}
\label{tab:gaussfit}
\begin{flushleft} {\it Notes.}
Column 1: galaxy merger name as listed in Table~\ref{tab:results}.  
Column 2: redshift of component. 
Column 3: full-width-at-half-maximum of component. 
Column 4: peak optical depth of component. 
\end{flushleft}
\end{table*}
\end{appendices}
\bsp
\label{lastpage}
\end{document}